\documentclass[preprint]{ptephy_v1}
\preprintnumber{2001.03281}

\usepackage{url} 
\usepackage{subcaption}

\begin{document}
\title{Design and performance of a high-pressure xenon gas TPC as a prototype for a large-scale neutrinoless double-beta decay search}

\author[1]{S.~Ban*}

\affil[1]{Department of Physics, Kyoto University, Kyoto 606-8502, Japan \email{bansei0526@scphys.kyoto-u.ac.jp}}
\affil[2]{Advanced Research Center for Beam Science, Kyoto University, Kyoto 611-0011, Japan}
\affil[3]{Yokohama National University, Faculty of Engineering, Yokohama 240-8501, Japan}
\affil[4]{Department of Physics, Kobe University, Kobe 657-8501, Japan}
\affil[5]{Institute of Particle and Nuclear Studies, High Energy Accelerator Research Organization (KEK), Tsukuba 305-0801, Japan}
\affil[6]{Kamioka Observatory, Institute for Cosmic Ray Research, the University of Tokyo, Hida 506-1205, Japan}
\affil[7]{Kavli Institute for the Physics and Mathematics of the Universe (WPI), the University of Tokyo, Kashiwa 277-8568, Japan}
\affil[8]{Research Center for Neutrino Science, Tohoku University, Sendai 980-8578, Japan}

\author[1]{M.~Hirose} 
\author[1]{A.~K.~Ichikawa}
\author[2]{Y.~Iwashita}  
\author[1]{T.~Kikawa}
\author[3]{A.~Minamino}  
\author[4]{K.~Miuchi} 
\author[5]{T.~Nakadaira}
\author[6,7]{Y.~Nakajima} 
\author[4]{K.~D.~Nakamura} 
\author[1]{K.~Z.~Nakamura} 
\author[1]{T.~Nakaya} 
\author[1]{S.~Obara\thanks{Current address: Frontier Research Institute for Interdisciplinary Sciences, Tohoku University, Sendai, 980-8578, Japan}}
\author[5]{K.~Sakashita}
\author[6,7]{H.~Sekiya} 
\author[1]{B.~Sugashima}
\author[1]{S.~Tanaka} 
\author[8]{K.~Ueshima} 
\author[1]{M.~Yoshida}
\begin{abstract}
A high-pressure xenon gas time projection chamber, with a unique cellular readout structure based on electroluminescence, has been developed for a large-scale neutrinoless double-beta decay search. 
In order to evaluate the detector performance and validate its design, 
a 180~L size prototype is being constructed and its commissioning with partial detector has been performed. 
The obtained energy resolution at 4.0~bar is 1.73 $\pm$ 0.07\% (FWHM) at 511~keV. 
The energy resolution at the $^{136}$Xe neutrinoless double-beta decay Q-value is estimated to be between 0.79 and 1.52\% (FWHM) by extrapolation.
Reconstructed event topologies show patterns peculiar to track end-point which can be used to distinguish $0\nu\beta\beta$ signals from gamma-ray backgrounds.
\end{abstract}

\subjectindex{Nxxxxxx}

\maketitle
\section{Introduction} 
Whether the neutrino is Majorana type or not is a crucial question for particle physics and cosmology. 
If the answer is ``yes'', neutrinos may have played an central role in creating matter-antimatter asymmetric universe via the Leptogenesis scenario~\cite{Fukugita:1986hr}. 
Extremely light neutrino masses may also be related to the Majorana nature (Seesaw mechanism~\cite{yanagida1979,gellman1979}).
Currently the most practical method to confirm that neutrinos are Majorana particles is to observe neutrinoless double-beta decay ($0\nu\beta\beta$ decay).
The strictest lower limit on the half-life of $0\nu\beta\beta$ decay in ${}^{136}$Xe was obtained by the KamLAND-Zen experiment to be $1.07\times 10^{26}$ years (90\%~C.L.)~\cite{PhysRevLett.117.082503}. 
Because its lifetime is expected to be very long, the search for $0\nu\beta\beta$ decay requires a ton-scale target mass, an ultra-low radioactive environment, and powerful background rejection. 
High energy resolution is especially essential to distinguish $0\nu\beta\beta$ decay from continuous backgrounds such as double-beta decay accompanying emission of neutrinos ($2\nu\beta\beta$ decay).
High-pressure xenon gas time projection chambers (TPCs) meet these requirements~\cite{Nygren_2018}.
The application of high-pressure xenon gas TPCs for $0\nu\beta\beta$ decay searches is being actively pursued by the NEXT~\cite{nexttdr} and PandaX-III experiments~\cite{Chen2017}. 
The former has demonstrated high energy resolution in a high-pressure cabin has TPC using electroluminescence (EL)~\cite{next2013_511keV, next2013_511keV_2, Renner_2018}, and the latter is developing a detector with good tracking capabilities using MicroMegas~\cite{WANG2019162439}.

We are also developing a high-pressure xenon gas TPC, AXEL (A Xenon ElectroLuminescence) for $0\nu\beta\beta$ decay searches. 
A unique feature of AXEL is its cellular readout scheme which also utilizes EL, called the electroluminescence light collection cell (ELCC).
By using the ELCC, the AXEL detector has the potential for both high energy resolution and scalability. 
The concept and a proof-of-principle of the ELCC are described in~\cite{BAN2017185}. 
In this paper, we describe design of a larger prototype with a 180~L volume and evaluate its performance. 

\section{Detector design and construction}
\begin{figure}[htbp]
	\centering
	\includegraphics[width=0.8\linewidth]{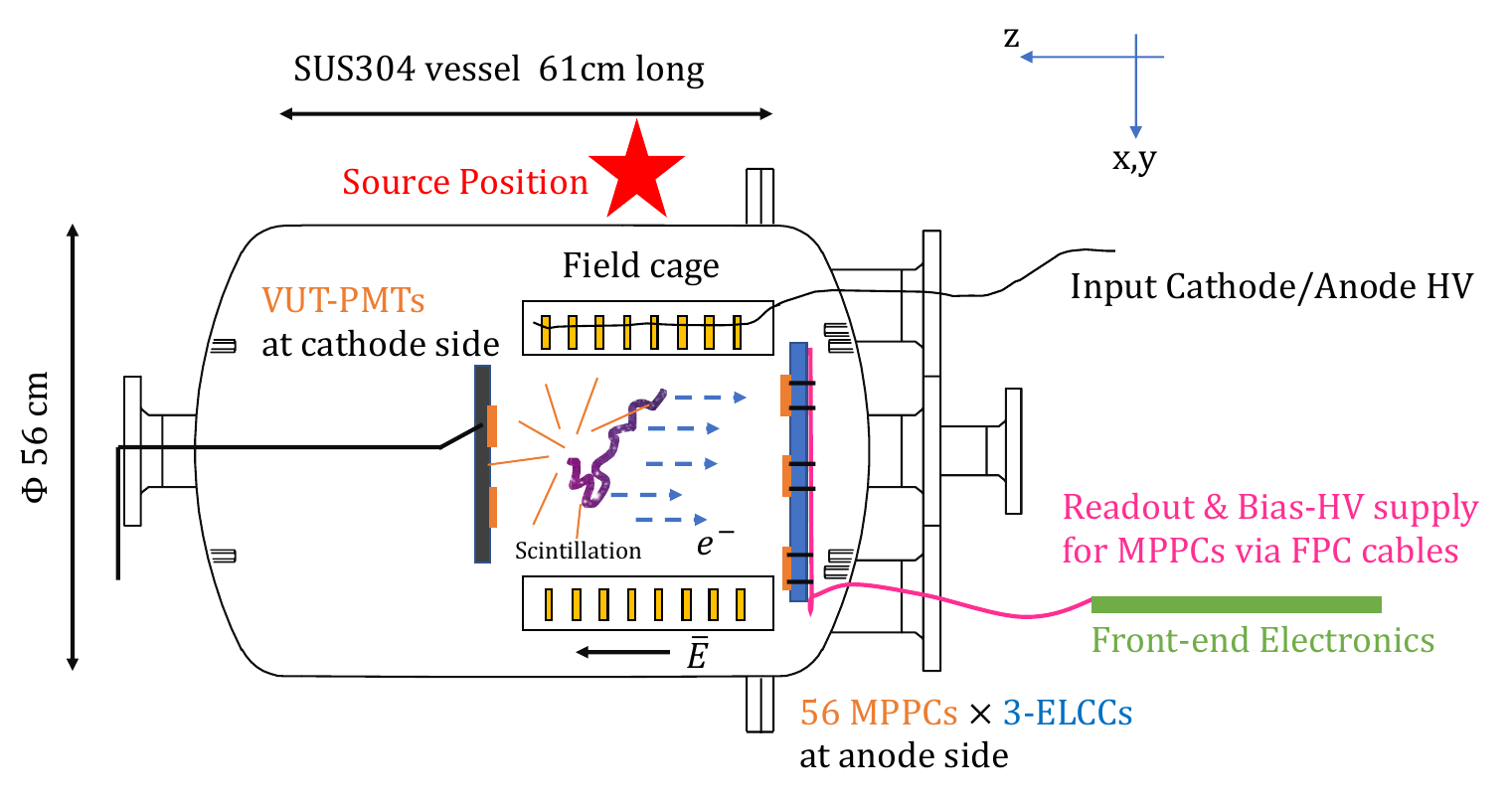} 
	\caption{Schematic view of the AXEL 180~L prototype chamber and a small field cage for the first phase run. Source position is indicated as a red star.} 
	\label{fig:detector} 
\end{figure}

The final goal of the 180~L size prototype is to evaluate the detector performance in the energy region around the $^{136}$Xe double-beta decay Q-value, 2458~keV. 
The detector components are housed in a vessel made of stainless steel (SUS304L) whose inner diameter is 547~mm, outer diameter is 559~mm, length is 610~mm, for a total volume of 180~L. 
The vessel can withstand up to 10~bar of pressure.
For the first phase of the 180~L prototype detector, we have constructed a small TPC whose size of the sensitive region is 15~cm diameter and 10~cm long, as shown in Figure~\ref{fig:detector}. 
The primary purpose of the first phase is an evaluation of the performance and validation of the design of the detector components, with 511~keV gamma-rays.
Ionization electrons are drifted to and detected by the ELCC (described in the next section) at the anode to measure the energy and topology of events in the volume.
Scintillation light is detected by PhotoMultiplier tubes (PMTs, R8520 Hamamatsu) at the cathode to determine the event timing.

\subsection{Electroluminescence Light Collection Cell}
\label{sec:ELCC}
\begin{figure}
    \centering
    \includegraphics[width=0.8\linewidth]{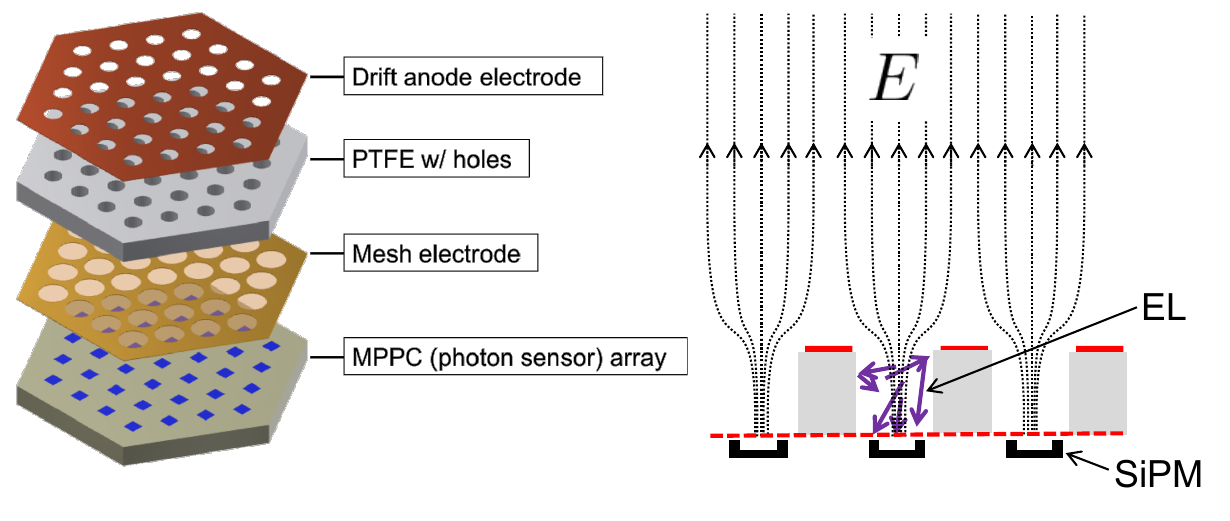}
    \caption{Expanded (left) and cross-sectional (right) views of the concept of ELCC. 
    The ELCC consists of four layers: drift anode electrode, PTFE plate, ground mesh electrode, and photon sensor array.
    Three top layers have patterned holes, and ionization electrons are drifted along the electric field into those cells.
    The EL photons are generated in the holes of the PTFE plate, between the anode electrode and the ground mesh electrode.
    }
    \label{fig:elcc_cross_sectional_view}
\end{figure}

The ELCC is a detector to readout ionization electron signals in the AXEL TPC~\cite{BAN2017185}.
Each cell is a pixel on an anode plane.
Ionization electrons are drawn into cells and produce EL photons that are detected by a Silicone Photomultiplier (SiPM) photon detector in each cell.
The EL process has less fluctuations than electron avalanche counterpart and it's therefore expected to have better energy resolution than detectors based on the electron avalanche~\cite{electroluminescence}.
The pixel structure enables tracking.
The ELCC plane consists of a drift anode electrode made of a 100~$\mu$m-thick copper plate with holes, a 5mm-thick polytetrafluoroethylene (PTFE) plate with holes, a ground potential (GND) mesh, and SiPMs as shown in Figure~\ref{fig:elcc_cross_sectional_view}.
The Hamamatsu $3 \times 3~{\rm mm}^2$ S13370 Multi-Pixel Photon Counter (MPPC), which is sensitive to the vacuum ultraviolet (VUV) EL produced in xenon, is used as the SiPM.
By applying a high voltage between the anode electrode and the GND mesh, an electric field that collects electrons is formed.
When the electric field exceeds the EL threshold, EL photons are generated.
The number of generated photons is given by the empirical formula~\cite{Monteiro_2007}
\begin{equation}
    Y_{\rm EL}/p = 140\,E/p-116,
    \label{eq:EL_field}
\end{equation}
where $Y_{\rm EL}$ is the photon yield for 1~cm electron drift, $E/p$ is the reduced electric field in units of kV/cm/bar and $p$ is the gas pressure in units of bar.

The dimension of the ELCC structure was optimized from the previous version~\cite{BAN2017185}.
In order to optimize the ELCC dimensions as follows, the energy resolution for 30~keV electrons was estimated for various configurations.
The electric field is calculated by using gmsh~\cite{gmsh} and Elmer~\cite{Elmer}.
Simulated electrons are generated 2~cm above the ELCC plane and tracked by Garfield$++$~\cite{garfieldpp}.
Electroluminescence photons are generated based on the electric field along the electron track and Equation~(\ref{eq:EL_field}) and the number of photons detected by the MPPC is calculated.
The aperture ratio of the GND mesh (50\%), photon detection efficiency of the MPPC (30\%), distance between the GND mesh and MPPC (1~mm), and PTFE reflectivity (66\%~\cite{doi:10.1063/1.3318681}) are taken into account.
The ELCC response is obtained from this procedure.
Next, 30~keV electrons are generated in the detector volume using Geant4~\cite{AGOSTINELLI2003250} and ionization electrons are generated, while taking the W-value and fano factor into account.
The position and time at 2~cm above the ELCC plane after the drift are calculated based on diffusion constants estimated by MAGBOLTZ~\cite{MAGBOLTZ}.
The number of detected photons for the 30~keV electron events is obtained using the ELCC response from above.
The optimization is done at 30~keV because xenon has characteristic X-rays of that energy, making it straightforward to compare with data.
The range of a 30~keV electron is 0.64~mm and it's diffusion over a 10~cm drift is 3~mm in xenon at 8~bar.
They are smaller than the typical ELCC cell pitch and enable sensitivity to EL yield non-uniformity within the ELCC cell.
The required energy resolution of 30~keV is 4.5\% FWHM or less, to achieve 0.5\% FWHM in terms of Q value.

The factors considered for the optimization are the cell pitch $l_{\rm pitch}$ and hole diameter $d_{\rm hole}$.
The EL field strength and the thickness of the PTFE plate, are fixed at 3~kV/cm/bar and 5~mm, respectively.
These numbers have been determined to give sufficient EL gain without necessitating excessive high voltage.
Cells are aligned in a hexagonal pattern since the distance between them is shorter than the square pattern for the same aperture ratio as shown in Figure~\ref{fig:elcc_cross_sectional_view}.

\begin{figure}[tb]
	\centering
	\includegraphics[width=0.7\linewidth]{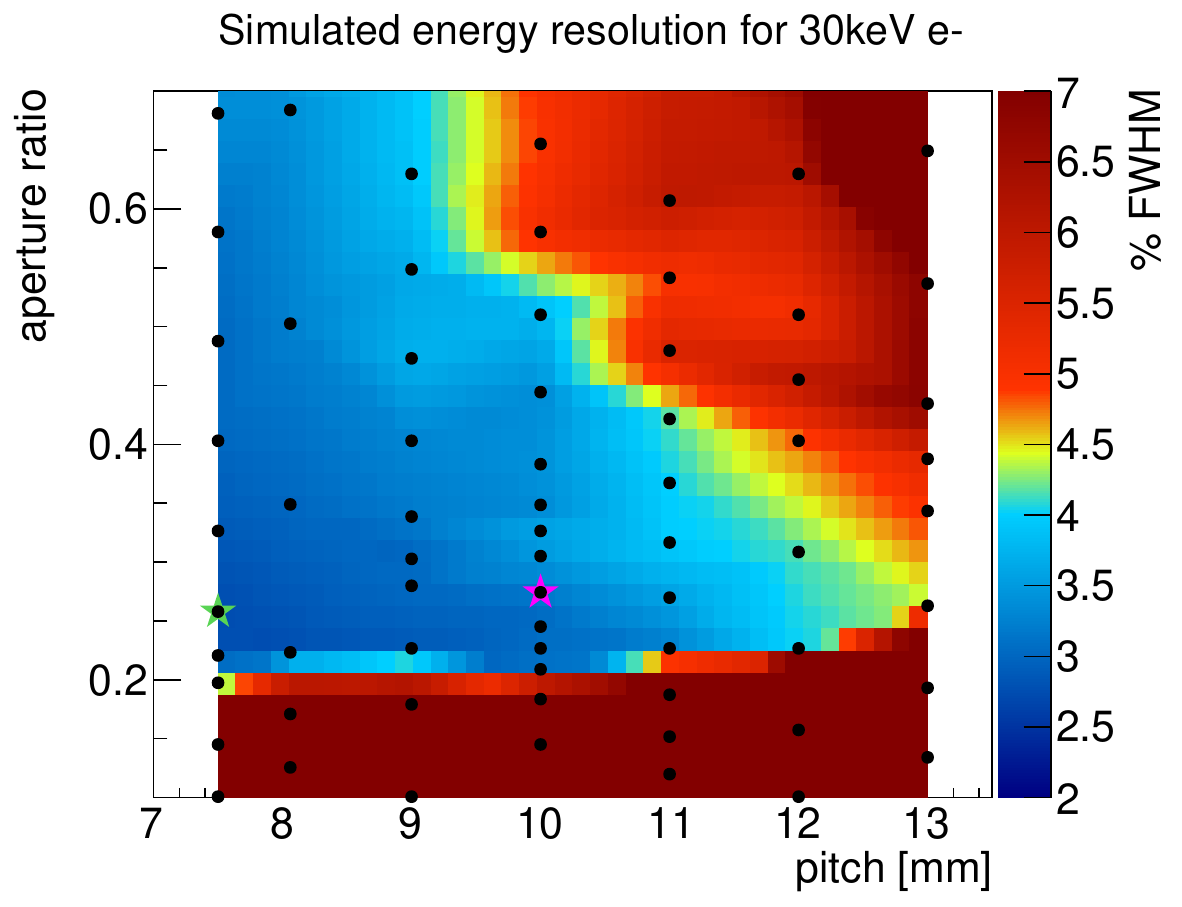} 
	\caption{
	    Expected energy resolution for 30~keV electrons, in xenon gas at 8~bar for various dimensions of the ELCC plane.
        Black dots represent simulation points.
        The color histogram is drawn by interpolation between the simulation points.
        The best parameters are found to be 7.5~mm for the pitch and 4~mm for the hole diameter and are shown by the green star.
        The adopted value, 10~mm for the cell pitch and 5.5~mm for the hole diameter, are shown by the magenta star.
        } 
	\label{fig:elcc_sim} 
\end{figure}

Figure~\ref{fig:elcc_sim} shows the expected energy resolution for 30~keV electrons in xenon gas at 8~bar for various cell pitches and anode aperture ratios.
If the aperture ratio is $0.2$ or less, the electron collection efficiency is poor, and the energy resolution deteriorates significantly.
For the aperture ratio larger than $0.2$, the smaller pitch and smaller aperture ratio give better energy resolution.
For tracking purposes, a cell pitch of 10~mm is fine enough because the typical diffusion of electrons for a 1~m drift is 1~cm.
A finer pitch increases the number of readout channels.
From Figure~\ref{fig:elcc_sim}, a 10~mm pitch has 4.5\% energy resolution, which is our requirement at 30~keV.
The hole diameter that minimizes the energy resolution is 5~mm, but 5.5~mm was adopted in consideration of machining accuracy.
Table~\ref{tab:pressure} summarizes the expected number of detected photons and energy resolution.
Since the measurement in this paper was performed at 4~bar, values at 4~bar are shown as well.

\begin{table}[bt]
    \centering
    \caption{Expected number of detected photons and energy resolution for 30~keV electrons with optimized dimensions: 10~mm pitch, 5.5~mm hole diameter and, 5~mm thick EL region.}
    \begin{tabular}{c c c} 
    Pressure & Number of photons & Energy resolution (FWHM)\\ \hline
    4~bar & 9100 & 3.4\% \\
    8~bar & 18000 & 3.2\% \\
    \end{tabular}
    \label{tab:pressure}
\end{table}

\begin{figure}[htb] 
	\centering
	\includegraphics[width=0.9\linewidth]{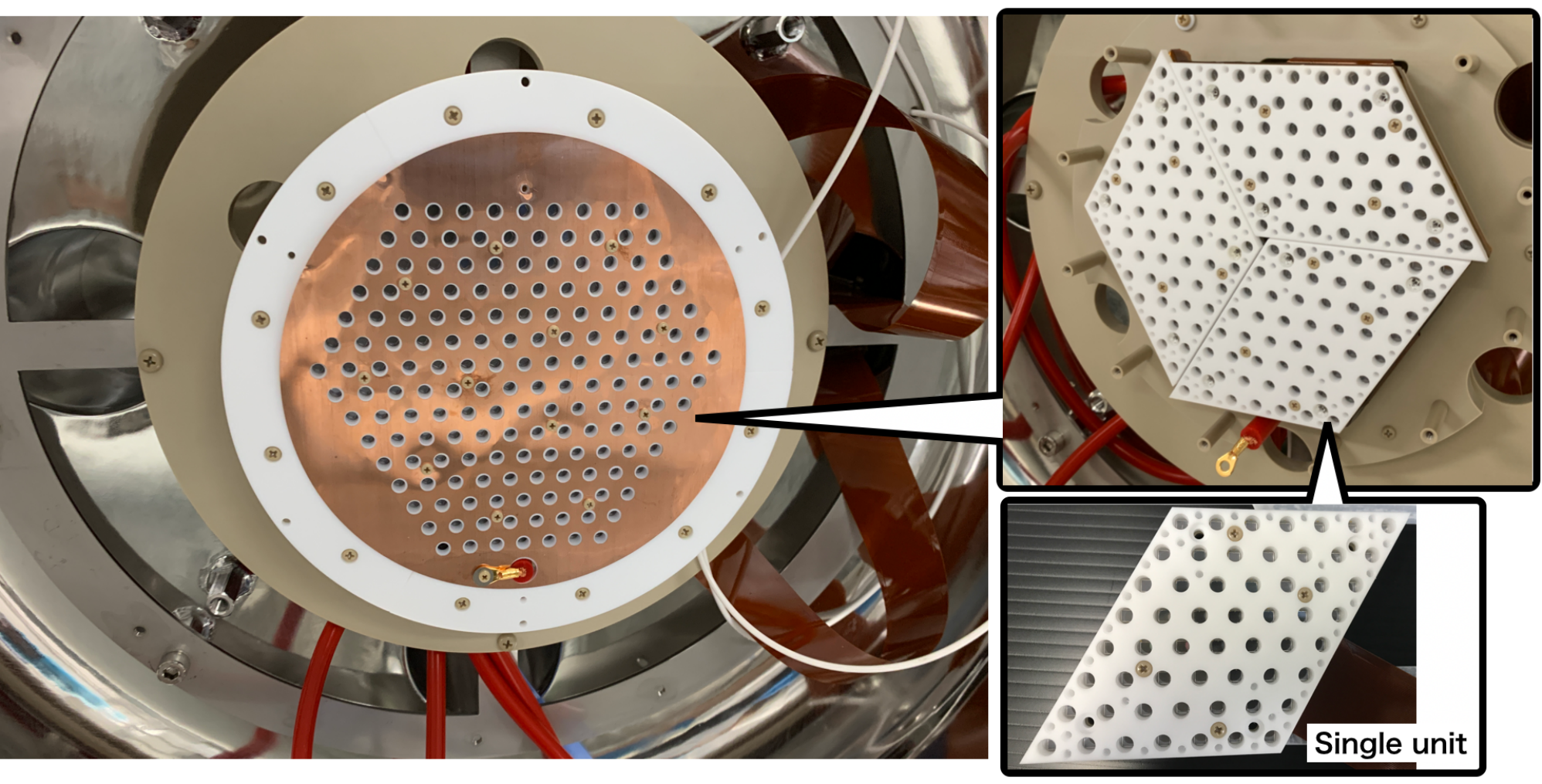}
	\caption{Photograph of the ELCC in the first phase of the 180~L prototype detector. The left picture shows the full ELCC including the anode electrode. The top right picture shows three ELCC units without the anode electrode. The bottom right picture shows a single unit of the ELCC.}
	\label{fig:ELCC_3units} 
\end{figure}

Following the result of the optimization, we constructed the ELCC with 10~mm pitch, 5.5~mm diameter anode electrode holes, a hexagonal cell pattern and a 5~mm thick EL region.
Figure~\ref{fig:ELCC_3units} shows the ELCC plane installed as the first phase 180~L prototype.
The ELCC plane of this first phase detector consists of three units.
It is extendable to a larger size by adding units.
Each ELCC unit has a trapezoidal shape and consists of a base plate made of polyetheretherketone (PEEK), MPPCs on the base, a PTFE body with cells, an anode electrode, ground electrode, and a flexible printed circuit (FPC) on which MPPC signal and bias lines are printed (see Section~\ref{sec:readout}).
A single unit has $7 \times 8$ channels. 
The total number of channels is 168.
The outer most 42~channels are set as veto channels and the remaining inner channels are regarded as fiducial channels.
The anode electrode is a single plate made of oxygen-free copper.
Tungsten mesh, whose wire diameter is 0.03~mm and aperture ratio is 78\%, is used as the GND mesh. 

\subsection{Drift electric field and field cage}
\label{sec:fieldcage}
To achieve high energy resolution, recombination between electrons and xenon ions should be suppressed because it reduces the number of initial ionization electrons and causes fluctuation in the signal size.
The rate of recombination is suppressed with a large drift field.
The large drift field is also preferred as it yields higher drift velocities and reduced diffusion.
In contrast, the efficiency of collecting ionization electrons into the ELCC decreases if the ratio of the intensity of the drift field to the EL field is not sufficiently low.
Hence we adopted a drift field of 100~V/cm/bar, which is acceptable for an EL field of 3~kV/cm/bar.
Non-uniformity of the drift field causes non-uniformity of the recombination rate.
Thus the drift field should be uniform to achieve high energy resolution.
Based on the results of previous studies on the relation between recombination and electric field~\cite{Serra_2015, Nakamura_2018}, we chose a target uniformity of $\pm 5\%$ for the intensity of the drift field.
Note that these previous studies are conducted with alpha particles from ${}^{222}$Rn in~\cite{Serra_2015} and ${}^{241}$Am in~\cite{Nakamura_2018}. The rate of recombination for ionization by alpha particles is higher than the one for ionization by electrons. Therefore this target value for uniformity is conservative.

The drift field is formed by a field cage which consists of a cathode mesh electrode on the PMT side, an anode electrode corresponding to the top electrode of ELCC, and ring electrodes aligned between the cathode and the anode.
The ring electrodes are band-shaped copper strips with two different radii, one radius for inner and one for outer strips.
A small overlap between the inner strips and the outer strips shields the effect of the vessel wall, and maintains uniformity of the drift field over a large volume inside the field cage.
The electrodes are supported by PTFE rings which also act as a reflector of VUV scintillation light to increase detection efficiency by the PMTs.

\begin{figure}[htb]
    \begin{subfigure}{0.38\columnwidth}
        \centering
        \includegraphics[width=0.9\hsize]{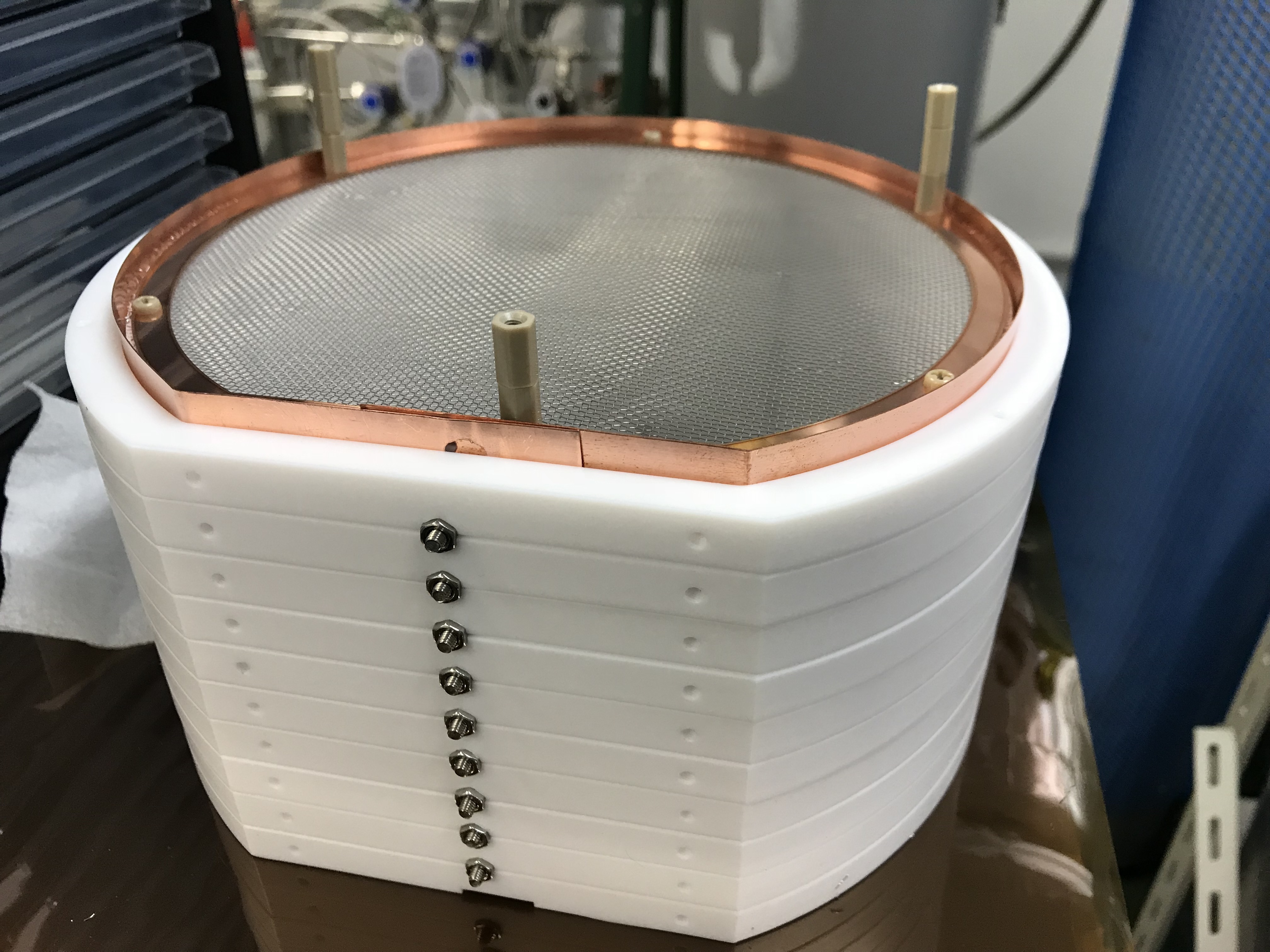}
        \caption{ }
        \label{fig:fieldcage_picture}
    \end{subfigure}
    \hspace{5mm}
    \begin{subfigure}{0.6\columnwidth}
        \centering
        \includegraphics[width=0.98\hsize]{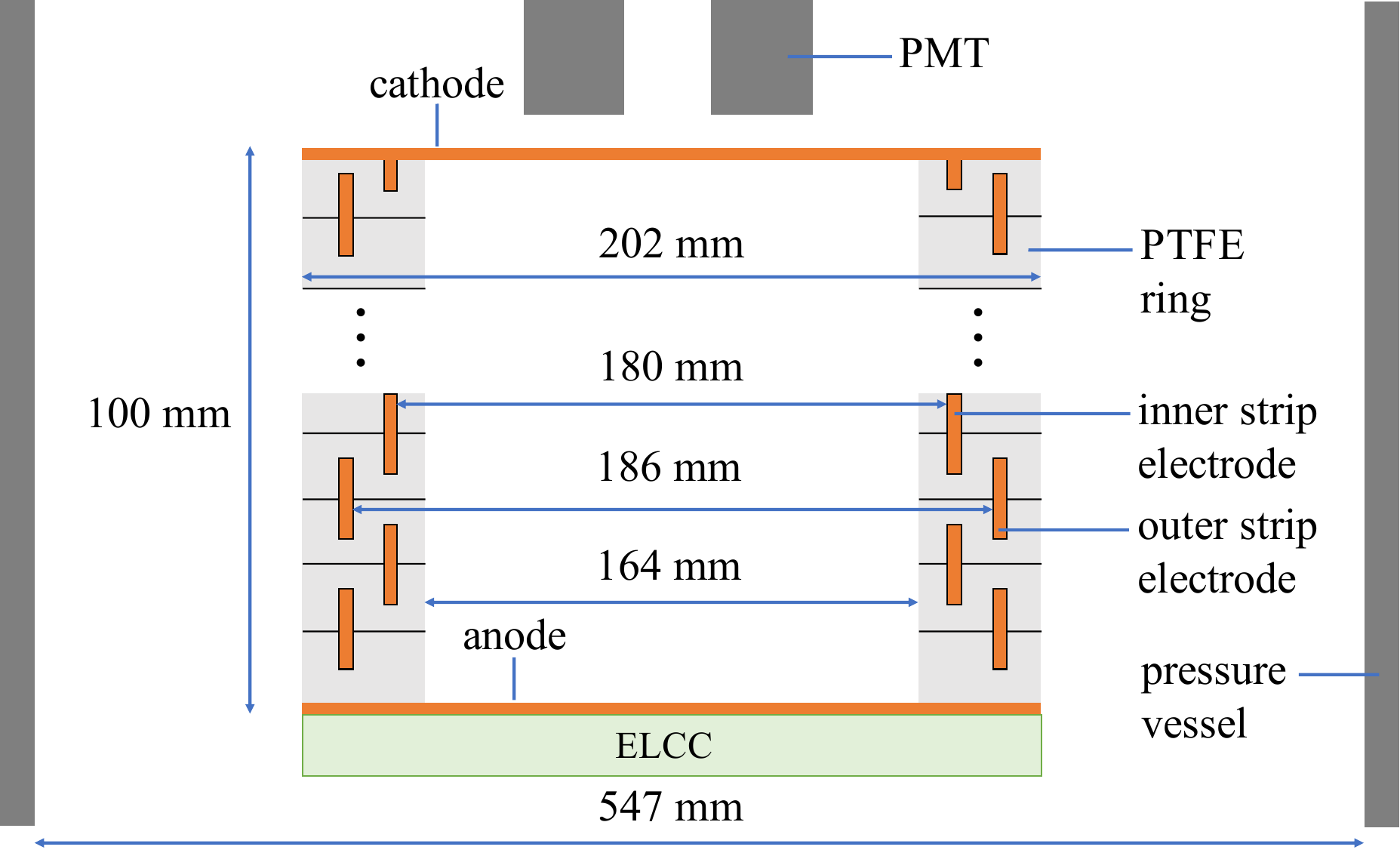}
        \caption{ }
        \label{fig:fieldcage_schematic}
    \end{subfigure}
    \caption{Field cage for the first phase of the 180~L prototype. The left is a photograph and the right is the schematic cross sectional view.}
\end{figure}

Figure~\ref{fig:fieldcage_picture} is a photograph of the field cage.
The thickness and the width of the strips are 0.3~mm and 12~mm, respectively.
Five inner electrodes and five outer electrodes are arranged at 10~mm intervals with 2~mm overlaps, resulting in a total drift length of 10~cm.
The cathode electrode is a stainless steel mesh. The wire diameter of the cathode mesh is $\phi$0.2~mm, and the wire is woven in an interval of 20~wires per one inch. Thus the aperture ratio of the cathode mesh is 71\%. The mesh used for the first phase prototype has a deflection, which changes the drift length. This deflection is roughly estimated to be within $\pm$1~cm.
The anode electrode and the strip electrodes are connected in serial by ten 100~M$\Omega$~resistors and the last inner electrode is connected directly to the cathode electrode.
The detailed dimensions are shown in Figure~\ref{fig:fieldcage_schematic}.

\begin{figure}[tb]
    \centering
    \includegraphics[width=0.85\hsize]{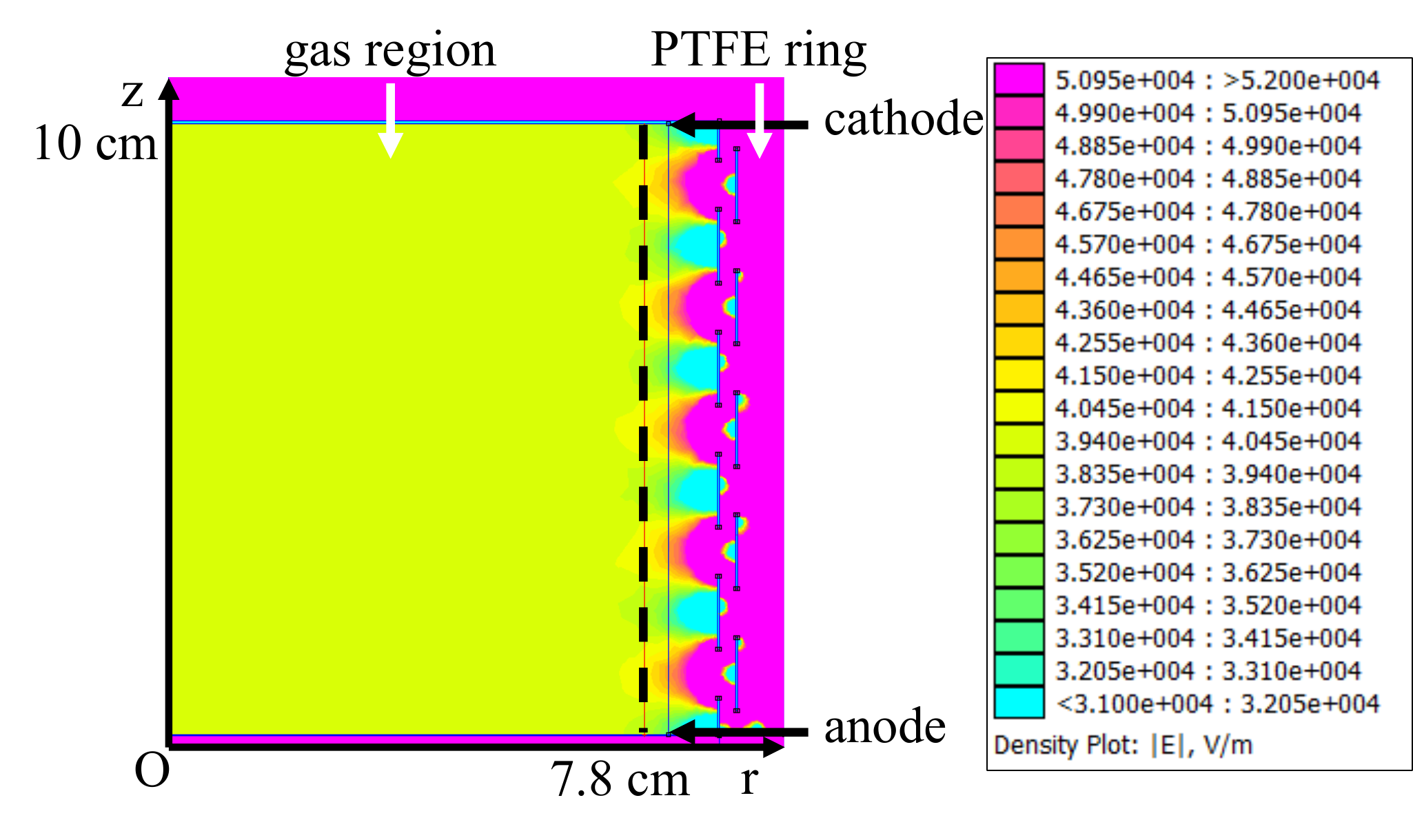}
    \caption{Field intensity plotted on the field cage geometry. The field intensity deviates from 100~V/cm/bar by $\pm$ 5\% where $r>7.8$~cm.}
    \label{fig:field_intensity}
\end{figure}

Figure~\ref{fig:field_intensity} depicts the electric field intensity calculated with FEMM~\cite{femm}.
The voltages were set to the values used in the measurement, $-6.0$~kV for the anode ($V_{\rm anode}$), $-10.0$~kV for the cathode ($V_{\rm cathode}$), and 0~V for the pressure vessel.
These values correspond to a 3~kV/cm/bar EL field and a 100~V/cm/bar drift field for xenon gas at 4.0~bar.
The result of the calculation shows that the requirement of 100~V/cm/bar~$\pm$~5\% is satisfied up to 4~mm inside the field cage ($r \leq 7.8$~cm) and covers the entire ELCC area.

\subsection{Signal readout}
\label{sec:readout} 
\begin{figure}
    \centering
    \begin{subfigure}{0.4\columnwidth}
        \centering
        \includegraphics[width=1.0\columnwidth]{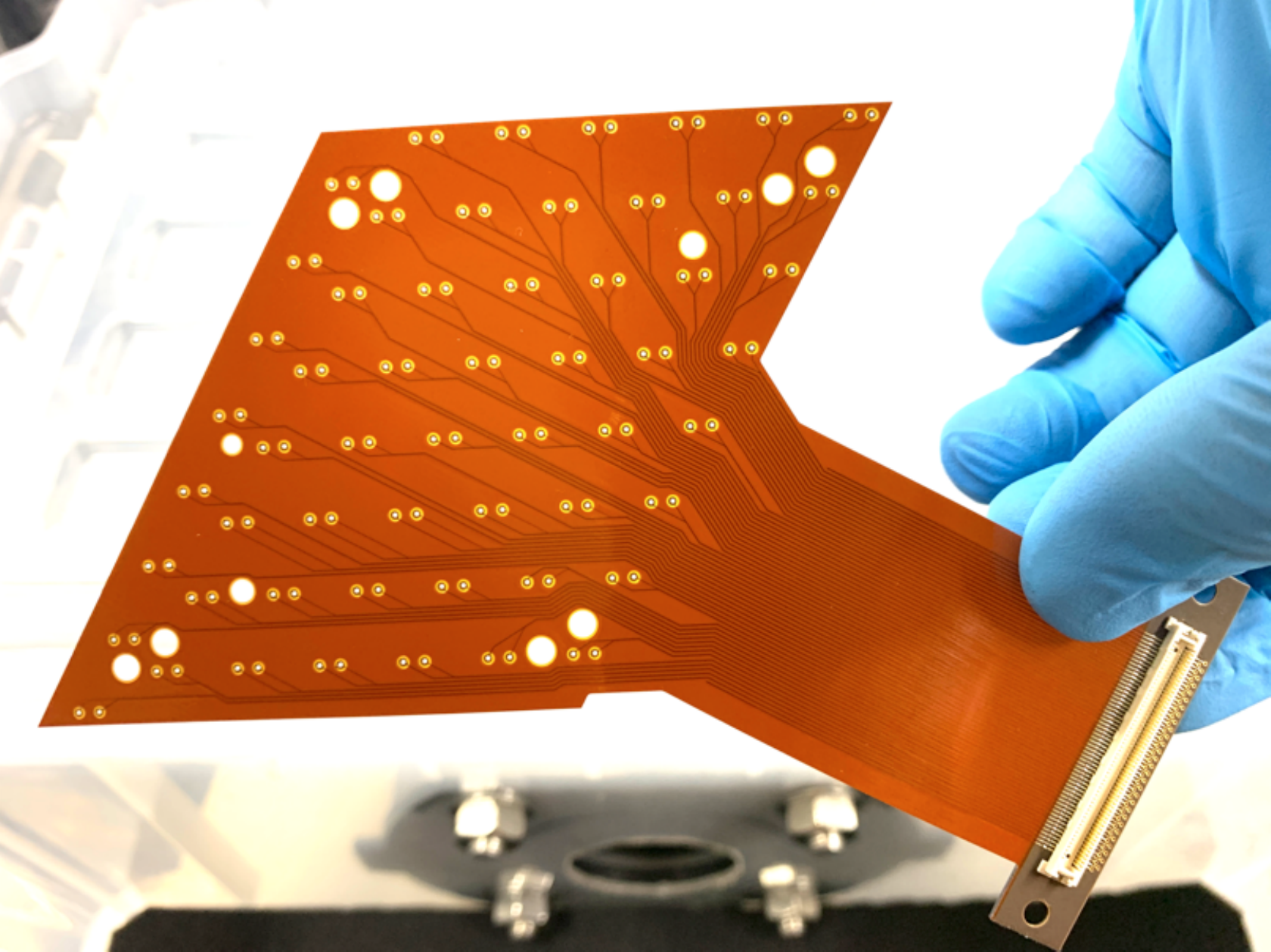}
        \caption{Unit FPC}
        \label{fig:unitFPC}
    \end{subfigure}
    \hspace{5mm}
     \begin{subfigure}{0.4\columnwidth}
        \centering
        \includegraphics[width=1.0\columnwidth]{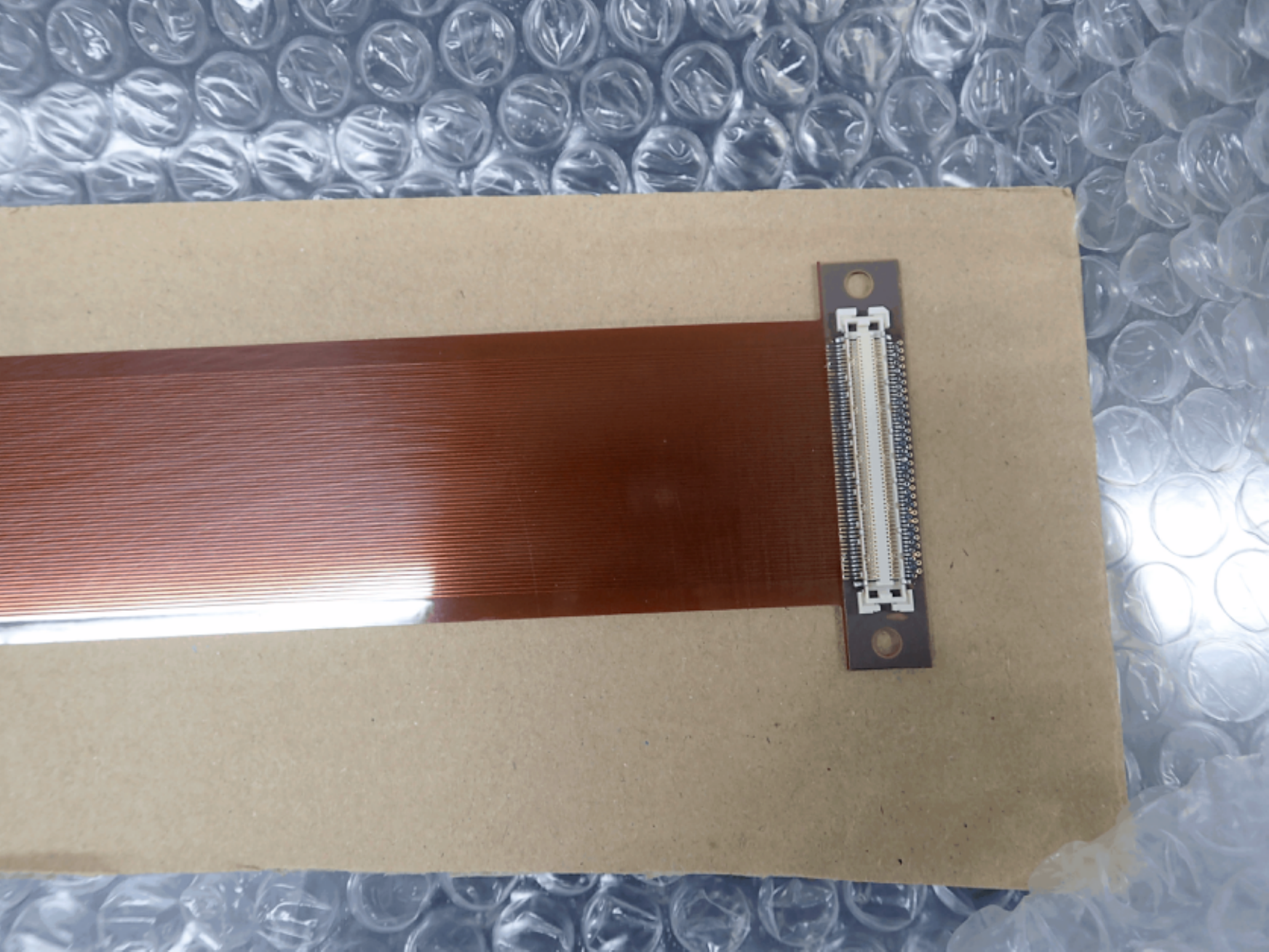}
        \caption{Cable FPC}
        \label{fig:cableFPC}
    \end{subfigure}
    \caption{FPCs for signal readout and application of bias voltage}
    \label{fig:FPC}
\end{figure}

At the bottom of the ELCC unit 56~MPPCs are mounted on another circuit FPC (Unit-FPC, see Figure~\ref{fig:unitFPC}) with connector pins.
The lengths of the bias and signal lines on the Unit-FPC are slightly different among MPPCs, but the timing differences are negligibly small comparing to the time scale of EL light emission.

In order to connect an ELCC unit with a front-end electronics board (FEB) through the 83.1~mm diameter feedthrough of the chamber, we chose a double-sided FPC-based cable, a picture of which is shown in Figure~\ref{fig:cableFPC}. 
Twelve cable FPCs are mounted collectively on a feedthrough flange with epoxy molding.
One FPC cable mounts 56~MPPC signal lines on the top side, 56~bias voltage lines on the bottom side, and four ground lines: 116~lines in total.
The FPC cable is 30~mm in width and 500~mm in length and consists of a coverlay (50~$\mu$m), an adhesive (35~$\mu$m), a copper trace (33~$\mu$m), an adhesive (20~$\mu$m), a base polyimide (25~$\mu$m), an adhesive (20~$\mu$m), a copper trace (33~$\mu$m), an adhesive (35~$\mu$m), and a coverlay (50~$\mu$m).
To suppress the cross-talk from neighboring lines, the signal and bias lines are designed to be 0.1~mm in width and 0.5~mm in pitch.
The basic design of this FPC was developed by the NEXT collaboration~\cite{nexttdr}. 
The Unit-FPC, the FPC cables, and the FEB are connected with FX11-LA connectors from Hirose electric.

We developed a dedicated FEB that has two types of ADCs for different amplifier gain to achieve a wide dynamic range from 1~photon to $\sim 10^4$~photons/$\mu$s. 
One $40~\mathrm{MS/s}$, $2~\mathrm{V_{pp}}$, $12~\mathrm{bit}$ ADC is connected to higher gain amplifier for every eight MPPCs via a multiplexer and is used for MPPC gain calibration.
The other is, a 5~MHz ADC and connected to a lower gain amplifier, is used for physics data taking.
The effective gain of this FEB is 0.2~pC per ADC count.
One FEB has 56~readout channels and acquires a waveform for up to $600$~$\mu$s.
This FEB also provides bias voltage to the MPPCs.
The voltage is adjustable for each MPPC.
Data are transferred to a DAQ PC via SiTCP Ethernet~\cite{weko_188_1}.
Details of the FEB is described in~\cite{kznamamura2019}.

\subsection{Gas system}
\begin{figure}[htb]
    \centering
    \includegraphics[width=0.9\linewidth]{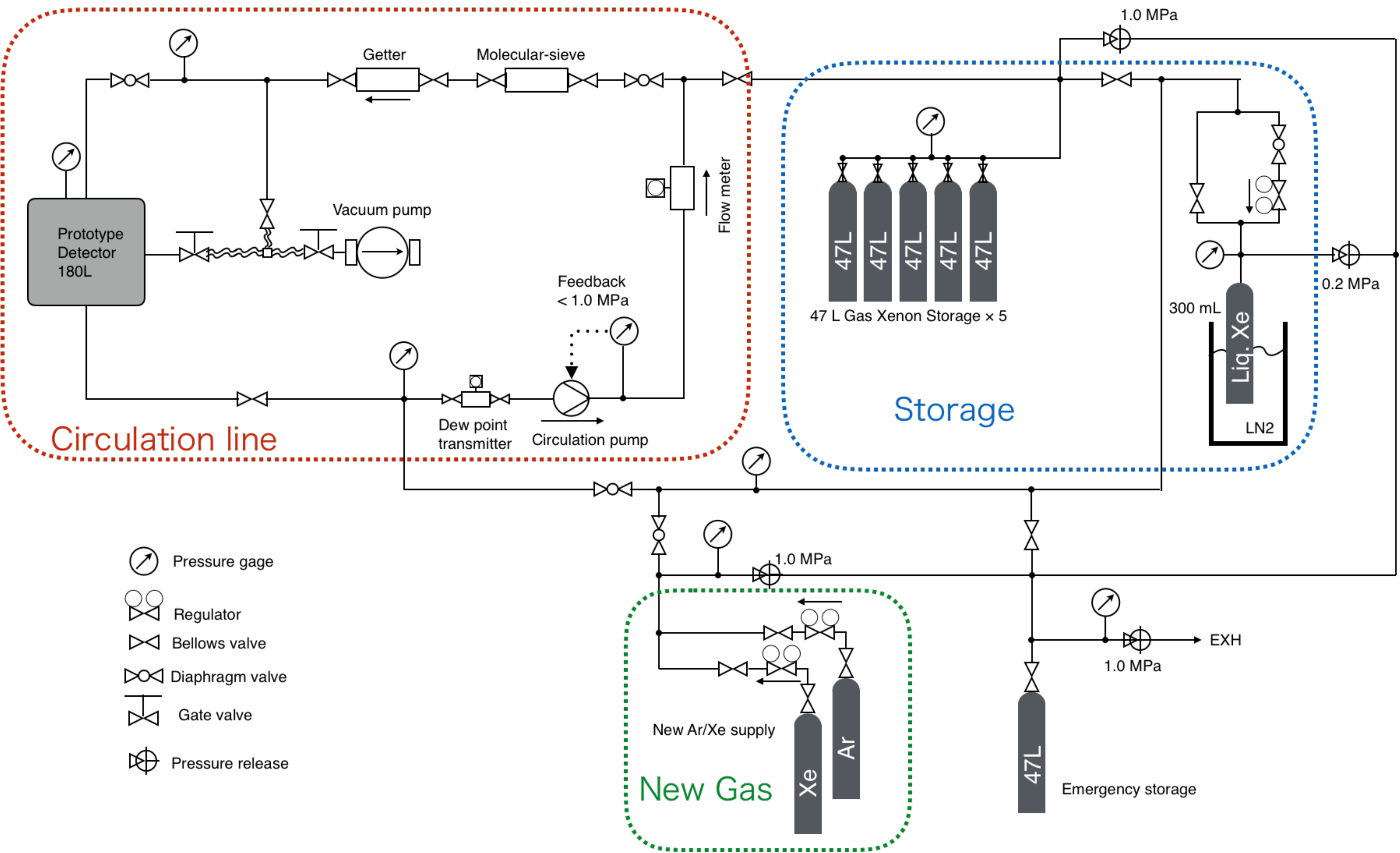}
    \caption{Schematic view of the gas system}
    \label{fig:gas_system}
\end{figure}

A diagram of the xenon gas system is shown in Figure~\ref{fig:gas_system}. 
It is equipped with a vacuum pump system, a circulation compressor (MB-601HPAL, IBS), purification systems, gas analyzers, gas storage, and the AXEL prototype detector.
Gas lines and the pressure vessel can store up to 10~bar of gas.
Before filling the detector, the vessel is purged with argon gas and exhausted to $10^{-2}$~Pa to reduce outgassing from the detector using a scroll pump (ISP-250C, ANEST IWATA) and a turbo-molecular pump (TG350FCAB, OSAKA VACUUM).
The rate of outgassing was $\sim 8.0\times10^{-5}$~Pa$\cdot$m$^3\cdot$s${}^{-1}$.
The xenon gas can be stored in five 47~L cylinders in the gaseous phase and in a 300~mL bin as liquid while the detector is opened.
The system can hold a total of 2100~normal liters of xenon gas.

For the measurement below, we used about 4~bar of natural xenon gas with less than 100~ppm of contaminants.
The gas is circulated during the data taking and a molecular-sieve (MC1-902FV, SAES) and a nitrogen getter (API-GETTER-I-RE, API) maintain the purity of xenon gas.
A dew point transmitter (PURA, MICHELL Instruments) monitors the water concentration.
Pressure gauges (ZT67, Nagano Keiki) measure the pressure with a precision of $\pm 0.6$~bar and monitor with much better resolution.

\section{Measurement} 
\label{sec:measurement}
\begin{figure}
    \centering
    \includegraphics[width=1.0\linewidth]{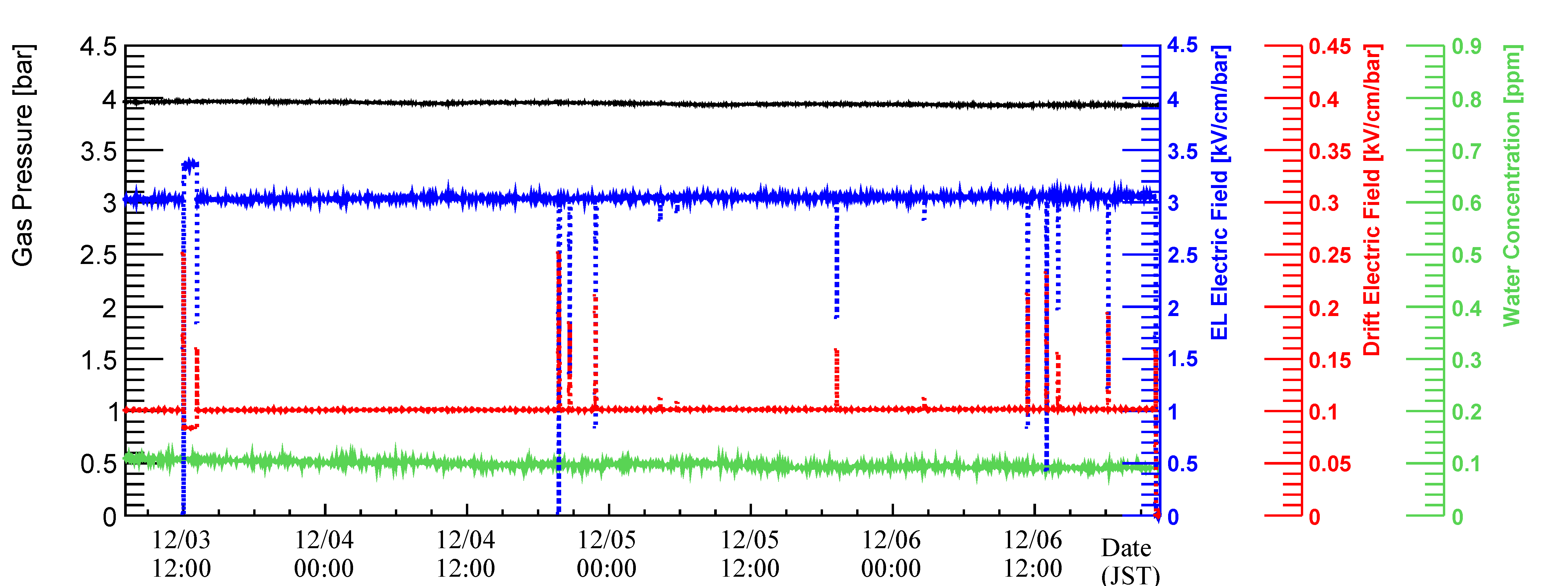}
    \caption{Monitor values during the data taking. The black line shows the gas pressure in the prototype detector, the blue line is EL electric field, the red line is drift electric field, and the green line is the water concentration.
    Note that no calibration was applied for these data so the absolute values have non-negligible systematic error of the pressure gauges and the dew point transmitter.
    Spikes in the electric field values are due to anode-voltage discharges and trips.}
    \label{fig:monitor}
\end{figure}

The detector performance was evaluated by irradiating it with 511~keV annihilation gamma-rays from a $^{22}$Na source.
As a first long term operation, we conducted this measurement at 4.0~bar, at which the high voltages are lower and commissioning is easier compared to the goal pressure of 8.0~bar.
It also enables a comparison with the previous measurement with the smaller prototype~\cite{BAN2017185}.
The intensity of the $^{22}$Na source was $7\times 10^5$~Bq and was set outside the vessel (see Figure~\ref{fig:detector}).
Data were taken for 4~days in December 2019.
Figure~\ref{fig:monitor} shows the various monitor data trends during the data taking.
The electric fields in the EL and drift regions were set to 3.0~kV/cm/bar and 100~V/cm/bar, respectively.
Although discharges happened mostly between the GND mesh and the anode electrode of the ELCC once per 6~hours on, shown as spikes in the figure, an interlock system cut and reset high voltage immediately.
Xenon gas was circulated at 10~L/min and purified by the molecular-sieve and the getter was operated at 400$^\circ$C.
The xenon gas pressure was stable at $4.0~\pm~0.6$~bar.
The water concentration was slightly modified by the purification but its variation was smaller than the systematic error of the dew point transmitter: $0.1~\pm~0.1$~ppm.
Except for the discharges, the detector had been stable for the entire data taking period.

The trigger was designed to issue when the height of the waveform sum of the inner channels exceeds a threshold and veto channels have no hits. However, due to a bug in the firmware, there were a few channel mis-identification between fiducial channels and veto channels. Complete veto was applied in the analysis stage, in stead.
In order to acquire 511~keV events efficiently, the threshold value was set high, roughly corresponding to 130~keV. A low-threshold trigger was set to acquire K$_\alpha$ (29.78~keV) events to calibrate the EL gain of each channel, as described in Section~\ref{sec:ELgain_correction}. 
The low-threshold trigger is reduced to 1/100 in order not to dominate the trigger rate.
Coincidence of two PMTs that are mounted on the cathode side is required in order to prevent contamination of accidental backgrounds.

In total, 8,100,166~events were acquired.
Of these, 1,000,000~events were used as a sample data set to determine fiducial cuts criteria and to establish the correction methods described below. The detector performance was evaluated using the entire data set.

\section{Analysis} 

\begin{figure}[htb] 
	    \centering
    	\includegraphics[width=0.97\linewidth]{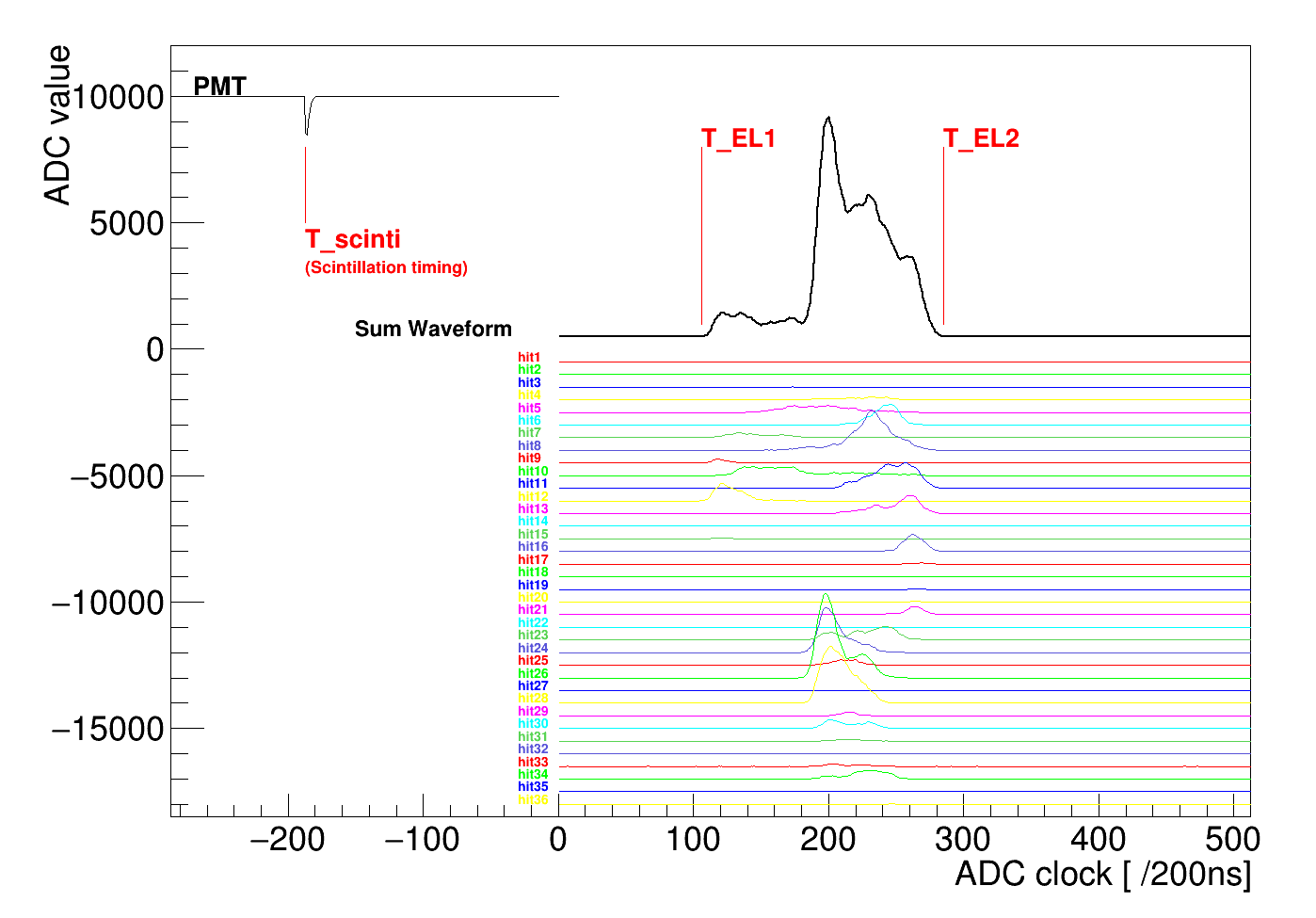} 
    	\caption{Typical waveform and definition of parameters. Sum waveform is the waveform sum of ELCC hit channels drawn as colored waveform.}
    	\label{fig:typical_waveform} 
\end{figure}

Typical signal waveforms of ELCC and PMTs are shown in Figure~\ref{fig:typical_waveform}.
Two PMTs at the cathode detect the xenon scintillation signal and 5--100~$\mu$s after that, EL signals are detected by the ELCC. 
The number of detected photons (photon counts) in each channel is obtained by integrating the waveform of each hit channel from the signal time of signal's rise to its fall and dividing it by the gain of the channel's MPPC. The MPPC gains are measured using dark current pulses as described in~\cite{BAN2017185}. For each hit channel, the non-linearity of the MPPC is corrected (Section~\ref{sec:MPPCcorrection}) and the EL gain is calibrated (Section~\ref{sec:ELgain_correction}).
The total number of photons in a given event is calculated by summing up the photon counts of all hit channels. The timing of the signal rise and fall of the event ($T_{\textrm{EL1}}$ and $T_{\textrm{EL2}}$) are defined as the earliest rise time and the latest fall time among the hit channels (see Figure~\ref{fig:typical_waveform}). 
The photon counts are then converted to deposited energy (see Section~\ref{sec:apply_cut_and_correction}).
The hit position along the drift direction ($z$-position) is reconstructed from the time interval between the PMT signal ($T_\textrm{Scinti}$) and the hit timing of the EL signal.

The detector performance was evaluated using photo-peaks at 511~keV (annihilation gamma-ray from $^{22}$Na), 29.78~keV (characteristic K$_\alpha$ X-ray), and 33.62~keV (characteristic K$_\beta$ X-ray).
To obtain clear photo-peaks, the fully-contained events in the fiducial region are selected (Section~\ref{sec:fiducial_cut}).
Additional corrections and cuts are described in  Section~\ref{sec:time_dependence}--\ref{sec:additional_z_cut}.

\subsection{MPPC non-linearity correction}
\label{sec:MPPCcorrection}

The linearity of the MPPCs degrades when the number of irradiated photons is comparable to the number of APD pixels constituting the MPPC.
This is because each APD pixel is operated in Geiger-mode and is not able to distinguish multiple photons.
Based on simulation, the maximum number of photons detected by a single MPPC is expected to reach $\sim 10^4$~photons in a few tens of $\mu$s for $0\nu\beta\beta$ signals.
Although $10^4$ is much more than the number of pixels of S13370~MPPC, $N_{\textrm{pixel}}=3600$, since the photons are distributed over tens of $\mu$s and the signal does not fully saturate and can be corrected.
The correction is performed with the following function,
\begin{equation} 
N_{\textrm{observed}}  = \frac{N'_{\textrm{observed}}}{1-\tau \cdot N'_{\textrm{observed}} /(N_{\textrm{pixel}}\cdot \Delta t)}, \label{eq:MPPCsatu}
\end{equation}
where $N'_{\textrm{observed}}$ is the number of observed photons before correction and $\Delta t$ is set to 200~ns, which corresponds to the sampling time of the 5~MS/s ADC.
This equation is derived in~\ref{apx:MPPCrecovery}.
Here $\tau$ is the MPPC pixel recovery time and was found to be around 120~ns according to our linearity measurement of MPPC. 
In this analysis, the same value is used for all MPPCs, 120~ns, as it gives the best energy resolution for the characteristic X-ray peaks ($\sim$ 30~keV) for the sample data set.

\subsection{EL gain calibration}
\label{sec:ELgain_correction}

\begin{figure}[htb] 
	    \centering
    	\includegraphics[width=0.3\linewidth]{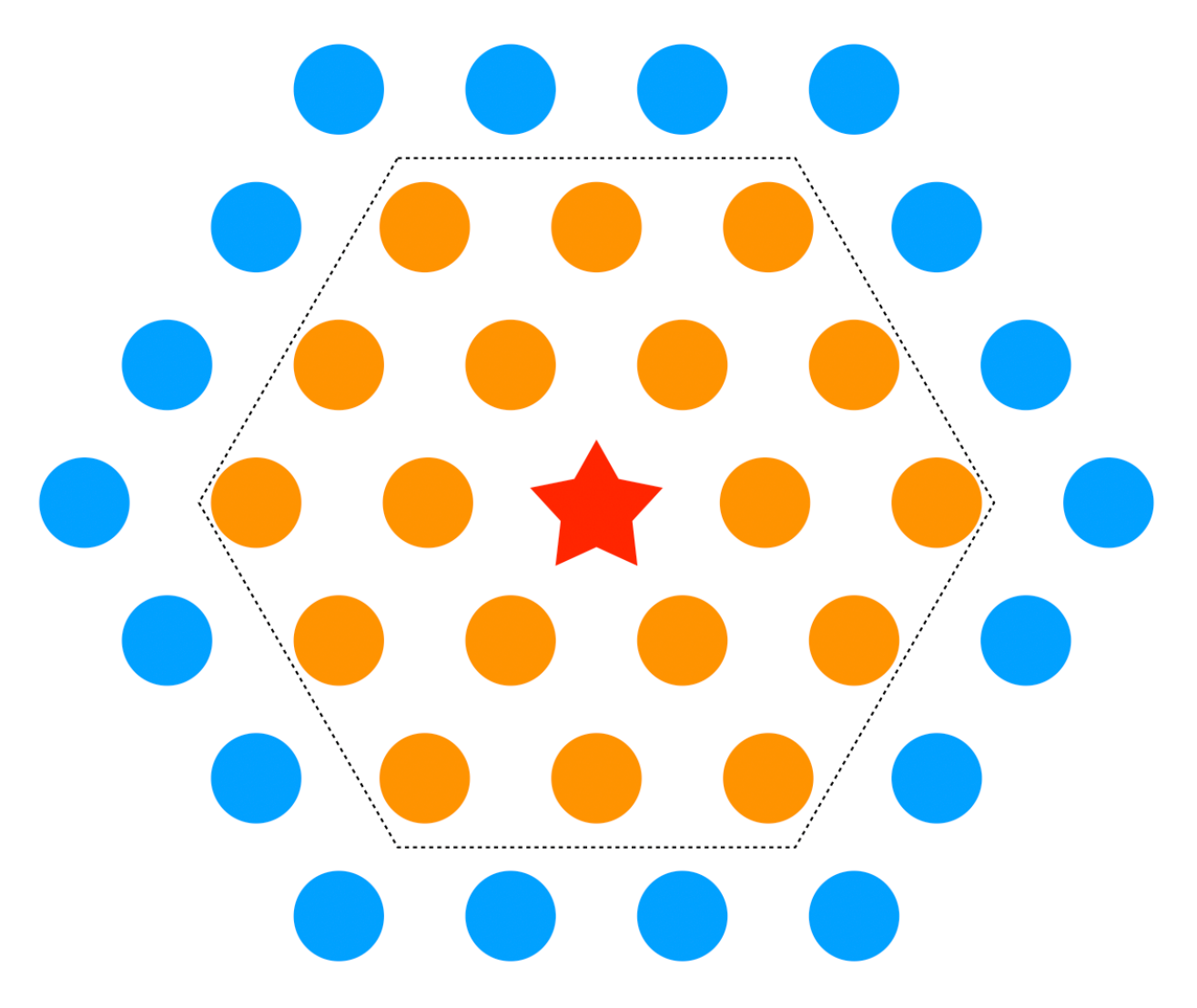} 
    	\caption{Selection of events for the EL gain calibration. The center channel represented as the red star is the channel being calibrated. That channel has to have the largest number of photons. All other channels except for the two layers of surrounding channels represented by orange circles are required not to have hits.}
    	\label{fig:ELgain_schematic} 
\end{figure}

Electroluminescence gain (EL gain) is defined as the average number of EL photons detected when one ionization electron enters in a cell and estimated for each cell using the photon counts of the K$_\alpha$ peak.
For each channel events are chosen in which that channel observed the largest number of photons, and there are no other hits except for the two layers of surrounding channels (see Figure~\ref{fig:ELgain_schematic}). 
The EL gain at 29.78~keV is obtained as the central value of the K$_\alpha$ peak after fitting with a Gaussian.

Throughout this process the gains of surrounding channels affect the calibration and therefore the calibration has to be iterated multiple times.
In this analysis, the EL gain calibration was repeated five times for all channels and additional four times for the fiducial channels.

\subsection{Fiducial volume cut}
\label{sec:fiducial_cut}

Events which only have hits in the fiducial channels of the ELCC plane are selected.

\begin{figure}[htb]
    \begin{subfigure}{0.49\columnwidth}
        \centering
        \includegraphics[width=0.98\hsize]{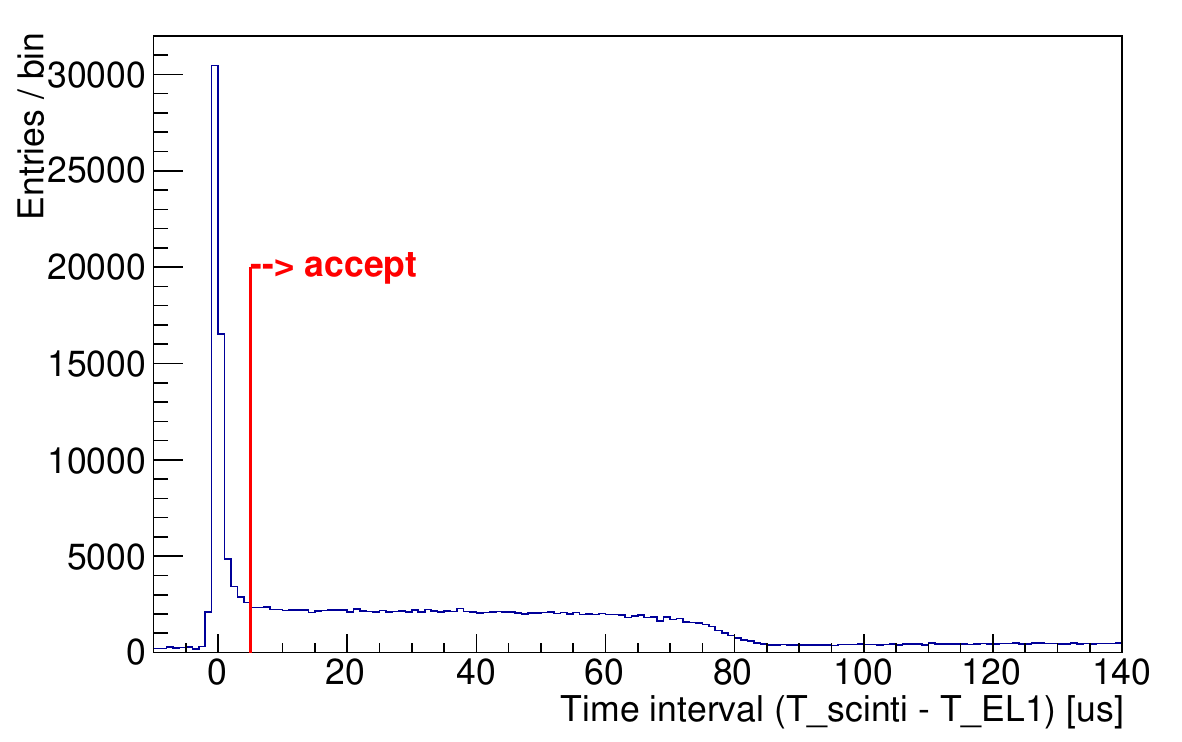}
        \caption{Time interval between $T_{\textrm{scinti}}$ and $T_{\textrm{EL1}}$}
        \label{fig:timeinterval_rise}
    \end{subfigure}
    \begin{subfigure}{0.49\columnwidth}
        \centering
        \includegraphics[width=0.98\hsize]{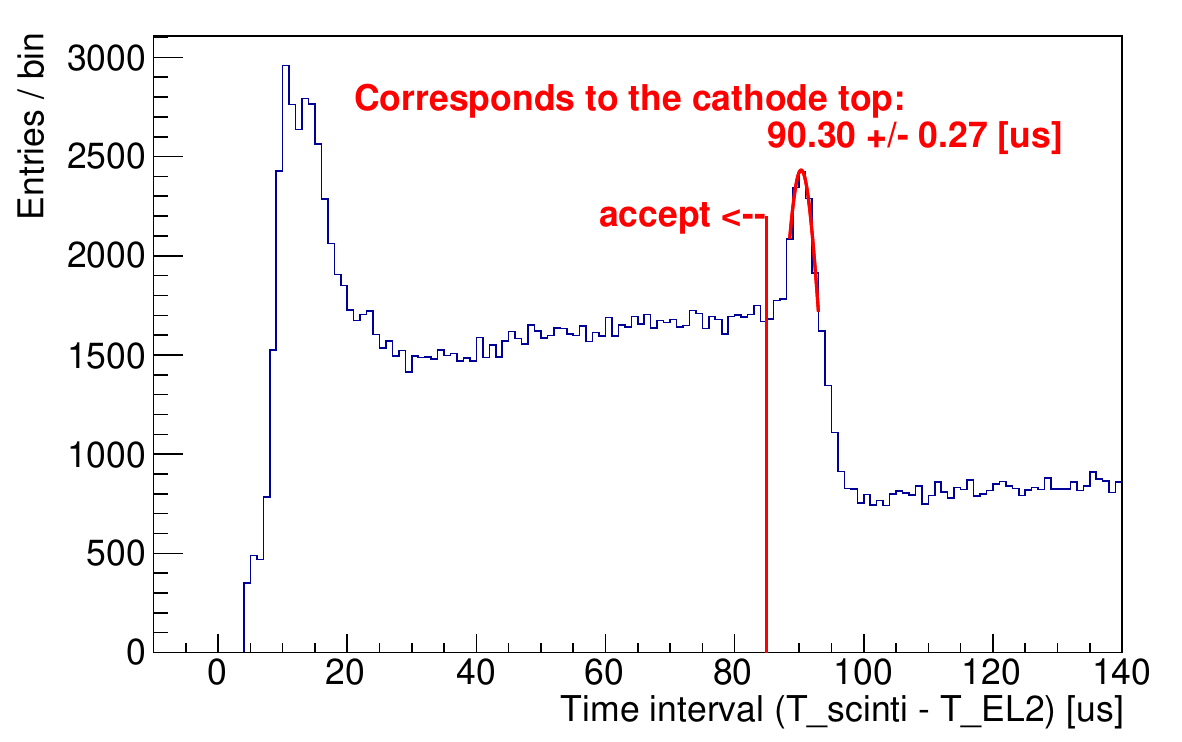}
        \caption{Time interval between $T_{\textrm{scinti}}$ and $T_{\textrm{EL2}}$}
        \label{fig:timeinterval_fall}
    \end{subfigure}
    \caption{Distributions of time interval between the scintillation signal and EL signal}
    \label{fig:timeinterval}
\end{figure}

Figure~\ref{fig:timeinterval} shows the distributions of the interval between $T_{\textrm{scinti}}$ and $T_{\textrm{EL1}}$ (Figure~\ref{fig:timeinterval_rise}) and between $T_{\textrm{scinti}}$ and $T_{\textrm{EL2}}$ (Figure~\ref{fig:timeinterval_fall}) of the sample data set after the fiducial channels cut.
The peak structure in Figure~\ref{fig:timeinterval_rise} corresponds to events which hit the anode electrode. The right peak in Figure~\ref{fig:timeinterval_fall} corresponds to the events which crossed the cathode electrode.
To chose fully-contained events along the drift direction, events whose time interval $T_{\textrm{scinti}}$ -- $T_{\textrm{EL1}}$ is more than 5.0~$\mu$s and time interval $T_{\textrm{scinti}}$ -- $T_{\textrm{EL2}}$ is less than 85~$\mu$s are selected.
The flat distribution above 100~$\mu$s in Figure~\ref{fig:timeinterval_fall} is due to
contamination of accidental hits and El signals. 
The contamination is high because the current PMT readout electronics has only timing information without waveform nor pulse height.

Since the cathode is at $z=$10 $\pm$ 1~cm the drift velocity of electrons in the detector can be measured by comparing the timing of events crossing the cathode with those at the anode ($z=$0~cm). The error of the cathode position comes from the distortion of the stainless mesh electrode as mentioned in Section~\ref{sec:fieldcage}. Fitting the cathode timing in Figure~\ref{fig:timeinterval_fall} with a Gaussian yields 90.30 $\pm$ 0.27~$\mu$s and thus the drift velocity is 0.11 $\pm$ 0.01~cm/$\mu$s.
This value is comparable to a previous study~\cite{next_2013_alpha}.
The $1\sigma$ peak width estimated from this fit is 4.54~$\mu$s and corresponds to 0.50~cm at the drift velocity of 0.11~cm/$\mu$s. This spread of the peak is caused by diffusion during drift and means that the reconstructed $z$ position has at most a 0.5~cm uncertainty.

\subsection{Time dependence correction}
\label{sec:time_dependence}
\begin{figure}[htb]
    \begin{subfigure}{0.49\columnwidth}
        \centering
        \includegraphics[width=0.99\hsize]{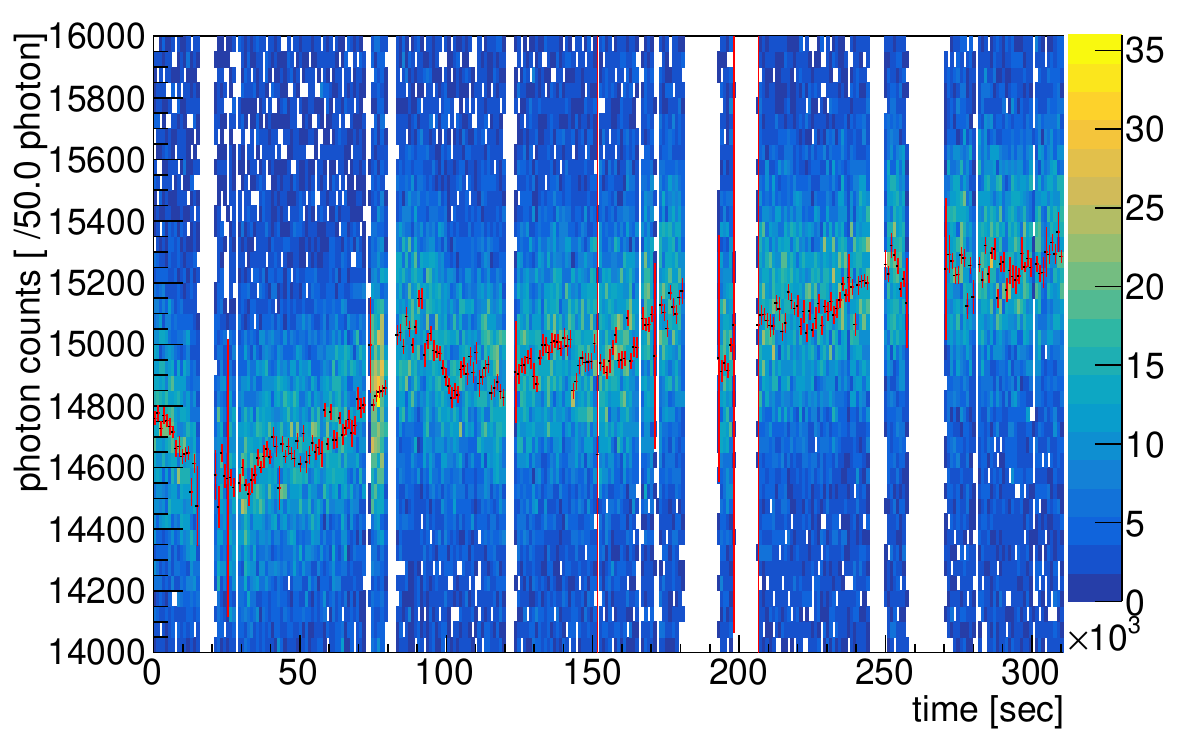}
        \caption{Before correction}
        \label{fig:timedependence_before}
    \end{subfigure}
    \hspace{1mm}
    \begin{subfigure}{0.49\columnwidth}
        \centering
        \includegraphics[width=0.99\hsize]{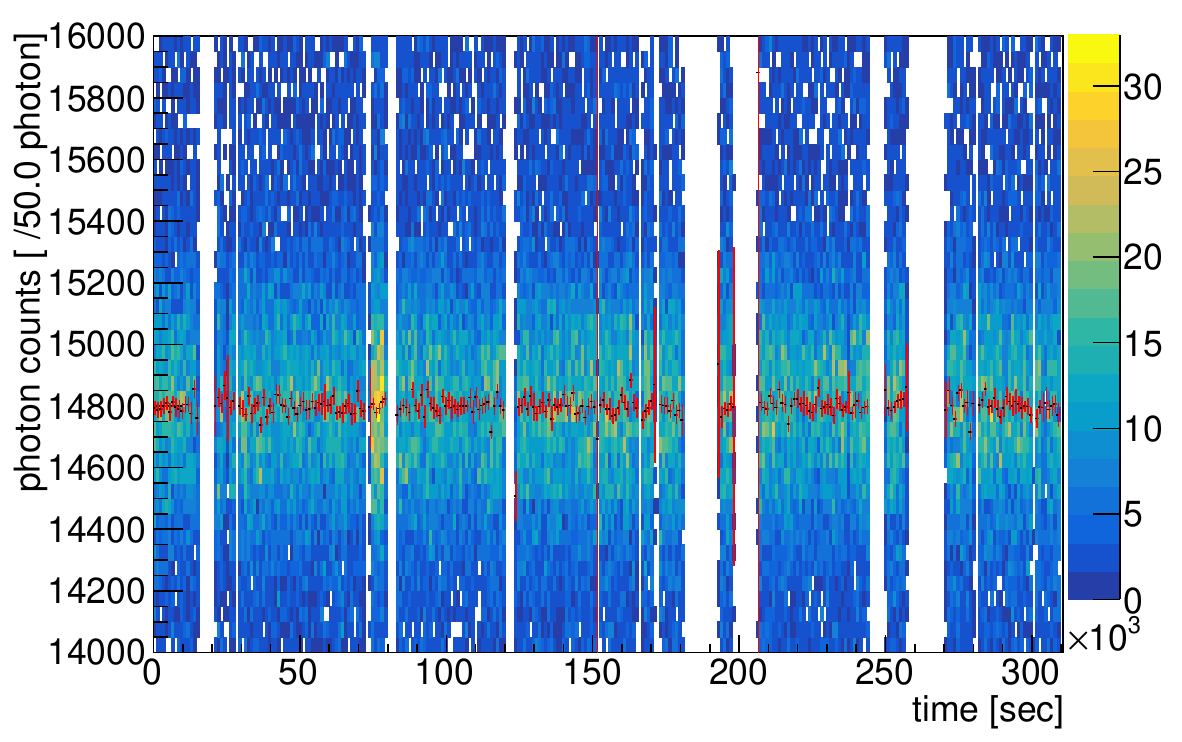}
        \caption{After correction}
        \label{fig:timedependence_after}
    \end{subfigure}
    \caption{Variation of light yield as a function of time before (a) and after (b) the correction. Black dots and red lines represent the K$_\alpha$ peak position and its fitting error in each time bin, respectively. Empty bins are run changes or periods of DAQ troubles.}
    \label{fig:timedependence}
\end{figure}

Figure~\ref{fig:timedependence_before} shows the time dependence of the light yield. 
The change of the light yield is possibly caused by an improvement of gas purity and change of the gas density.
The data acquisition period is divided into 300~bins, and correction for the time dependence of K$_\alpha$ peak to be flat is applied (Figure~\ref{fig:timedependence_after}).

\subsection{Correction for $z$-dependence}
\label{sec:z_dependence}
\begin{figure}[htb] 
	    \centering
    	\includegraphics[width=0.7\linewidth]{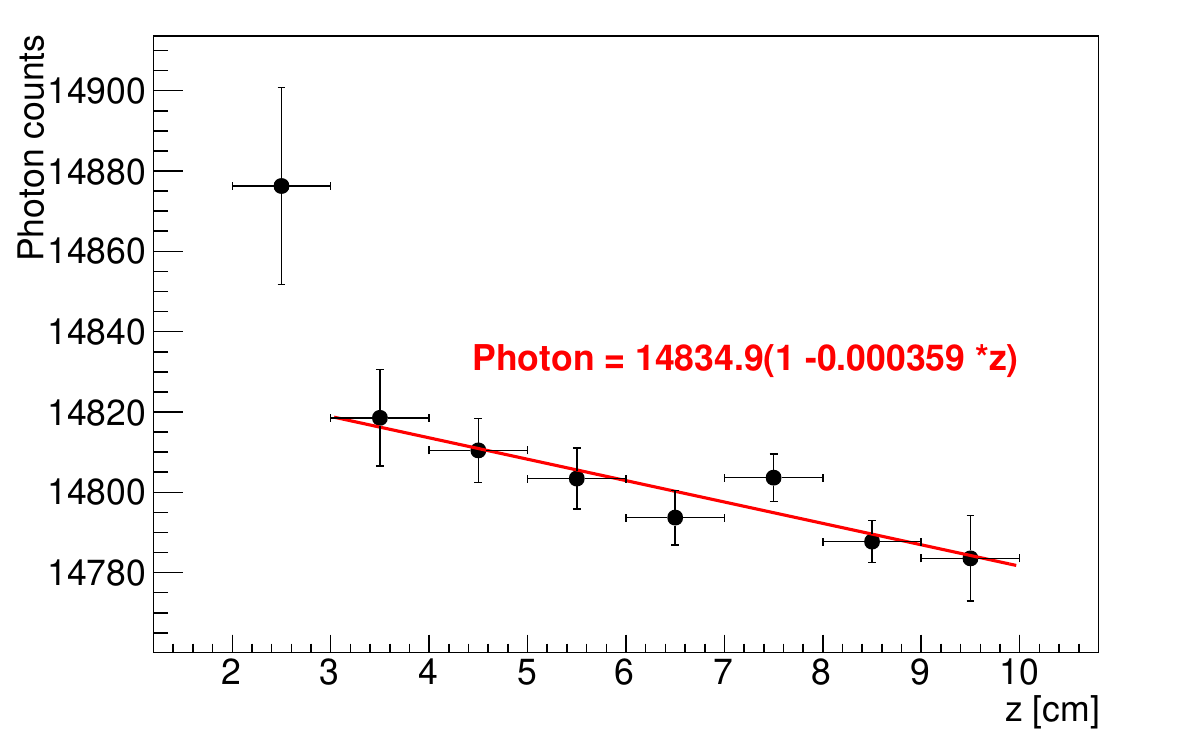} 
    	\caption{$z$-dependence of light yield for the K$_\alpha$ peak}
    	\label{fig:z_dependence} 
\end{figure}

Figure~\ref{fig:z_dependence} shows the photon counts of the K$_\alpha$ peak as a function of the $z$-position defined as the weighted average of the light amount.
The light yield decreases for events far from the ELCC. 
This is considered to be due to loss of ionization electrons due to capture by impurities such as oxygen. 
In the region below 3~cm, the light yield increases non-linearity. Non-uniformity of the light yield depending on the event position relative to the cell position is also observed in that region. 
The position dependence on the initial electron position 2~cm above the ELCC is also reproduced by a simulation at 4~bar.

The reduced yield in 3~cm$\leq z \leq$10~cm is fitted with a linear function. 
Using the fitted parameters, the $z$-dependence is corrected to be flat for every sampling point of the 5~MS$/$s ADC.

\subsection{Additional $z$ cut}
\label{sec:additional_z_cut}
As mentioned in Section~\ref{sec:z_dependence}, non-uniformity depending on the position relative to the cell position is remnant at $z<$3~cm. 
Therefore, events whose $z$-position at the time of their signal’s rise is less than 3.5~cm are cut.

\subsection{Result of cuts and corrections}
\label{sec:apply_cut_and_correction}
\begin{figure}[htb]
    \begin{subfigure}{0.49\columnwidth}
	    \centering
    	\includegraphics[width=0.99\hsize]{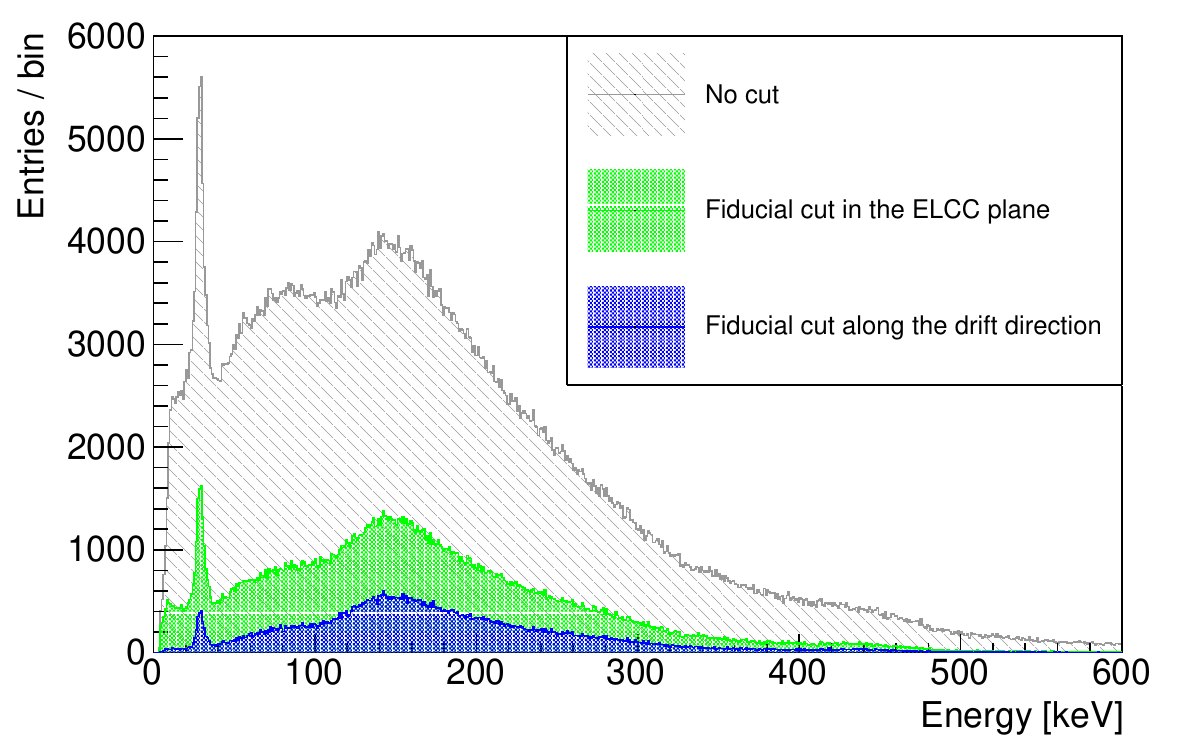}
    	\label{fig:cut_history_all} 
    \end{subfigure}
    \hspace{2mm}
    \begin{subfigure}{0.49\columnwidth}
        \centering
        \includegraphics[width=0.99\hsize]{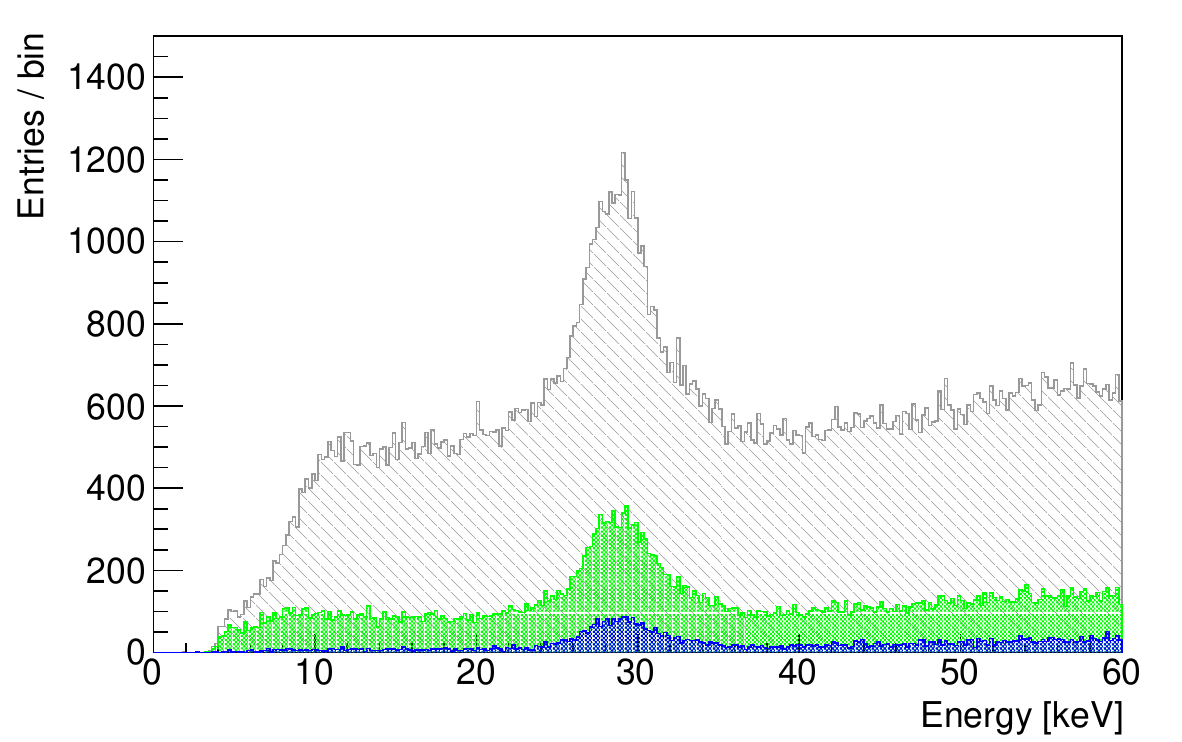}
        \label{fig:cut_history_low}
    \end{subfigure}
    \caption{Change of the energy spectrum after cuts for the sample data set. The right figure shows only the region around 30~keV.}
    \label{fig:cut_history}
\end{figure}

\begin{figure}[htb]
    \begin{subfigure}{0.49\columnwidth}
        \centering
        \includegraphics[width=1.0\hsize]{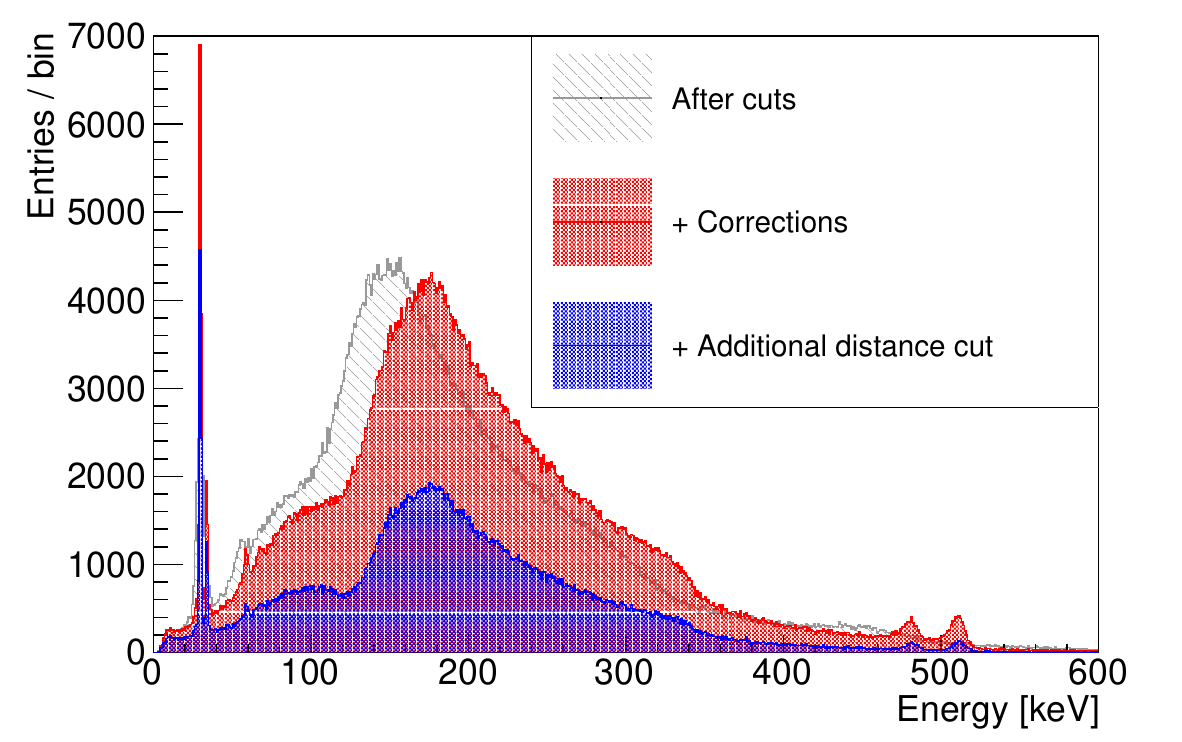}
        \label{fig:correction_history_all}
    \end{subfigure}
    \hspace{2mm}
    \begin{subfigure}{0.49\columnwidth}
        \centering
        \includegraphics[width=0.99\hsize]{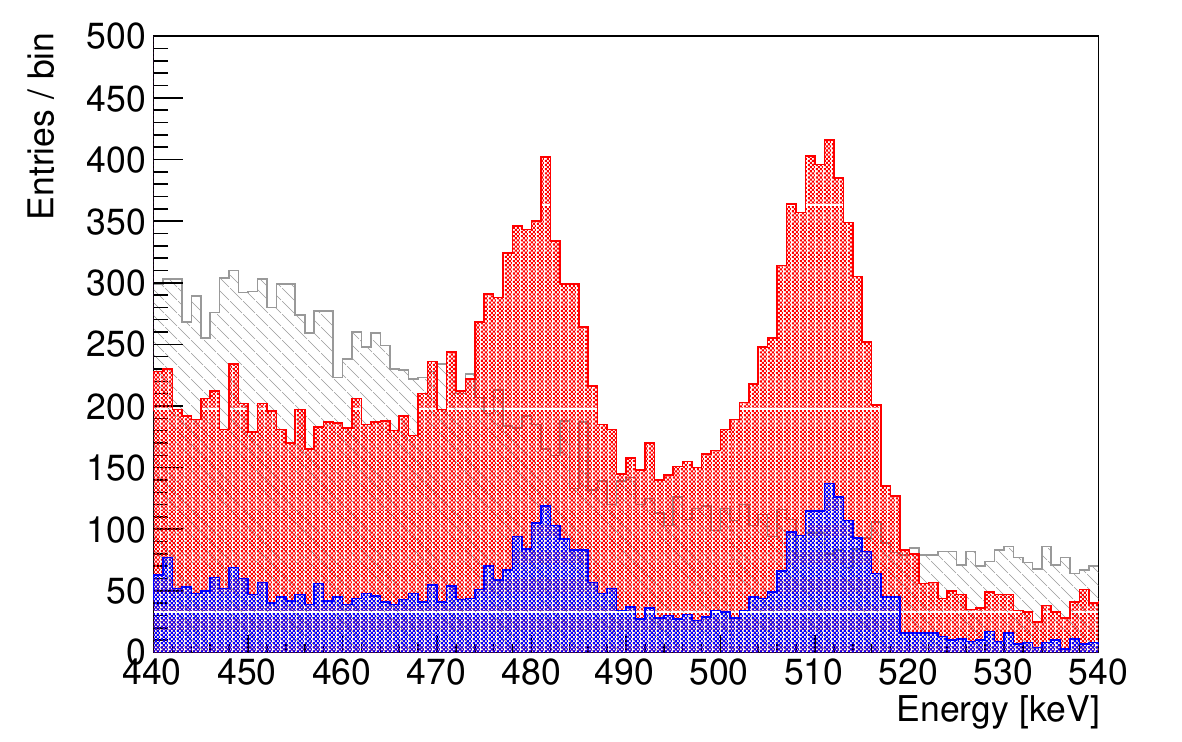}
        \label{fig:correction_history_high}
    \end{subfigure}
    \caption{Change of the energy spectrum by corrections and the additional distance cut for the whole data set. The right figure is the zoom up around 511~keV}
    \label{fig:correction_history}
\end{figure}

The change of the energy spectrum after each fiducial volume cut is shown for the sample data set in Figure~\ref{fig:cut_history}. 
Figure~\ref{fig:correction_history} shows the change in the energy spectrum after all corrections and the additional $z$ cut for the whole data set. 
After these corrections and cuts, peak structures at 511~keV and $\sim$480~keV (escape peaks) are clearly seen.
In these histograms, the energy scale is calibrated using the photon counts of two characteristic X-ray peaks (29.78~keV, 33.62~keV) and a 511~keV peak. 

\section{Detector Performance} 
\subsection{EL yield}
The total photon counts of K$_\alpha$ events and 511~keV events are $14805\pm 3.08$ and $256773\pm 140.3$, respectively.
The expected number of photons of 30~keV events at 4~bar was 9100 as mentioned in Section~\ref{sec:ELCC}.
This inconsistency may due to systematic error of MPPC detection efficiency, the reflect index of PTFE, and the angular dependence of incident photons relative to GND mesh.

\subsection{Electron lifetime}
The lifetime of electrons during drift is estimated from the $z$-dependence described in Section~\ref{sec:z_dependence}. 
The correction coefficient, $1-az$, can be cast as
\begin{equation}
    1-az \simeq \exp( -az ),
\end{equation}
where $a$ is the slope of the correction, $0.000359 \pm 0.000120$.
This leads to a $1/e$ decay length of
\begin{equation}
    1/( 0.000359 \pm 0.000120 ) = 2785.51 \pm 931.09\  \textrm{cm}. 
\end{equation}
Conversion to the electron lifetime using the drift velocity, 0.11 $\pm$ 0.01~cm/$\mu$s yields $25.32 \pm 8.77$~ms.

\subsection{Energy resolution}
\begin{figure}[htb]
        \centering
        \includegraphics[width=0.8\hsize]{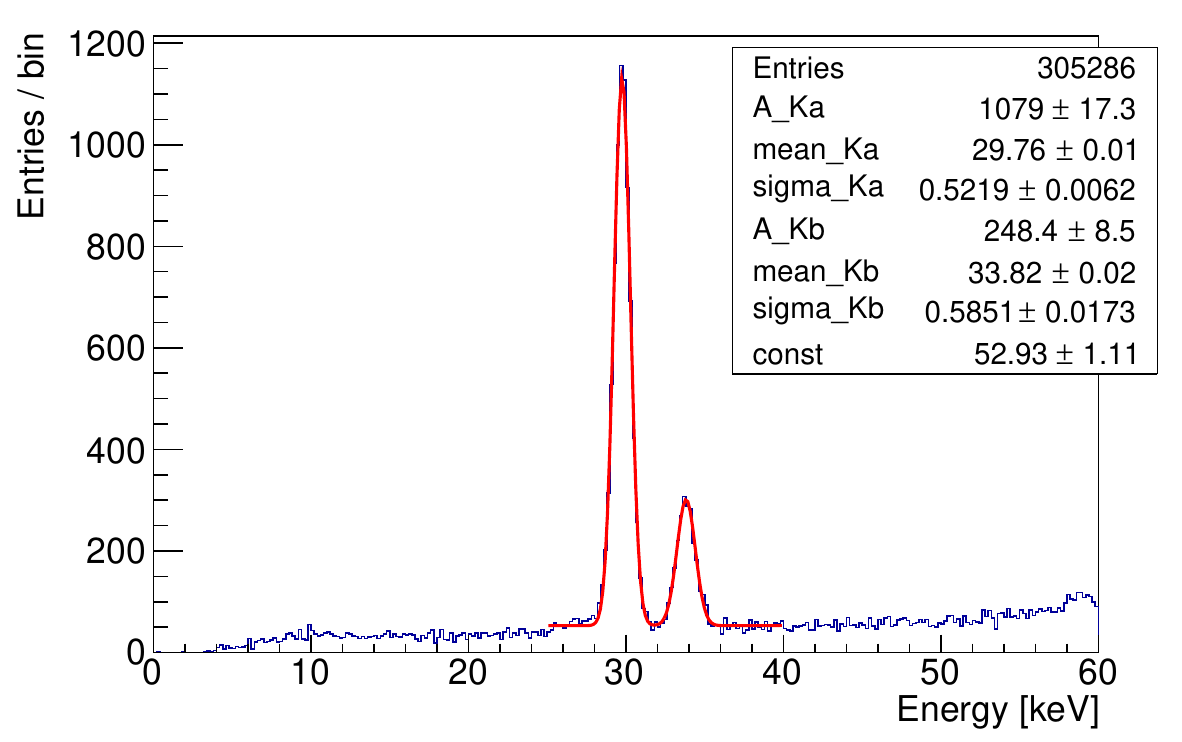}
        \caption{Energy spectrum around 30~keV and fit result to two Gaussian functions and a constant}
        \label{fig:fitting_30keV}
\end{figure}

\begin{figure}[htb]
        \centering
        \includegraphics[width=0.85\hsize]{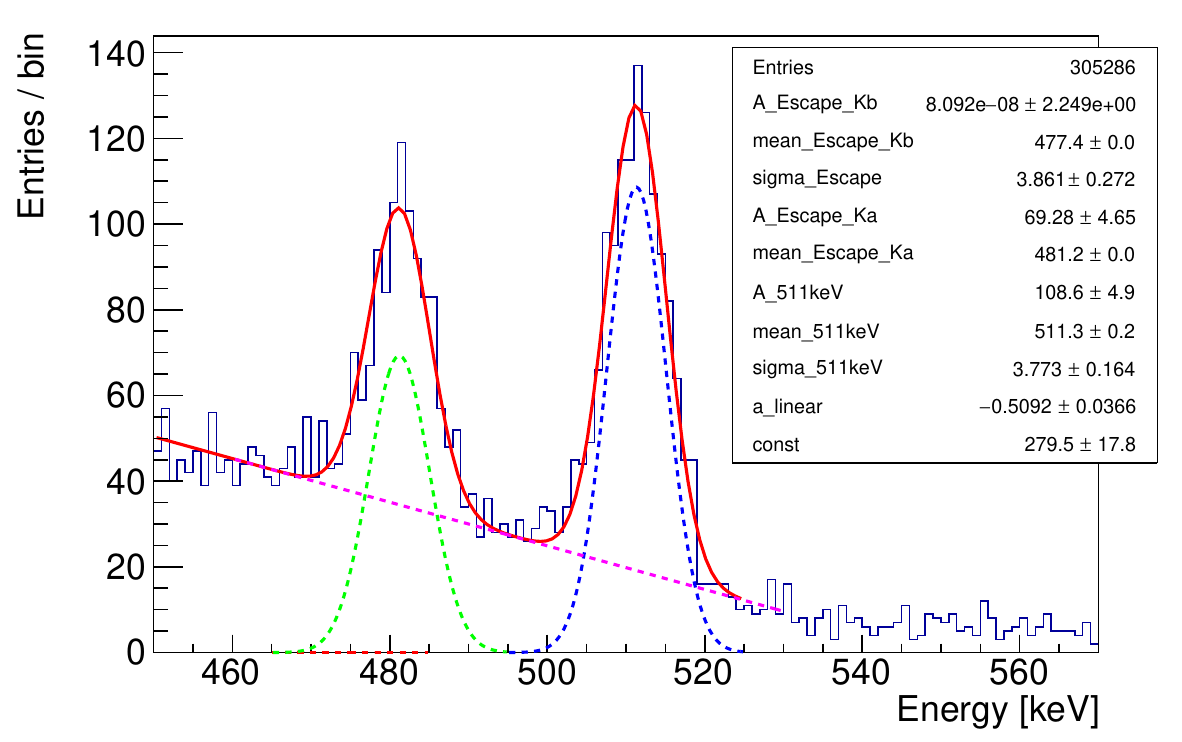}
        \caption{Energy spectrum around 511~keV and fit result. The peak structure at 480~keV consists of the K$_\alpha$ and K$_\beta$ escape peaks. The escape peaks are fitted with double-Gaussian and 511~keV peak is fitted with Gaussian. The continuum component is fitted with a linear function.}
        \label{fig:fitting_511keV}
\end{figure}

Figure~\ref{fig:fitting_30keV} shows the energy spectrum around the characteristic X-rays of xenon and their fitting result with double-Gaussian plus constant. 
The obtained energy resolutions are 4.10~$\pm$~0.05\% (FWHM) and 4.06~$\pm$~0.14\% (FWHM) for the K$_\alpha$ and K$_\beta$ peaks, respectively.
Figure~\ref{fig:fitting_511keV} shows the energy spectrum around the 480~keV and 511~keV. 
The peak at around 480~keV consists of the escape peaks of K$_\alpha$ (481.22~keV) and K$_\beta$ (477.38~keV). 
Accordingly, this peak was fitted using a double-Gaussian with the peak positions of the fitting function fixed to each characteristic energy. 
The 511~keV peak was fitted with single Gaussian.
A linear function was added to model continuum components.
The obtained energy resolution is 1.73~$\pm$~0.07\% (FWHM) for 511~keV, which corresponds to 0.79~$\pm$~0.03\% (FWHM) at the ${}^{136}$Xe $0\nu\beta\beta$ decay Q-value when extrapolated by $\sqrt{E}$.

The resolution at the Q-value was also evaluated assuming additional energy dependency term using the form $A\sqrt{E+BE^2}$, where $A$ and $B$ are the fitting parameters.
The resolutions at the K$_\alpha$, K$_\beta$, and 511~keV peaks are used to the evaluation with this function.

\begin{figure}[htb]
        \centering
        \includegraphics[width=0.8\hsize]{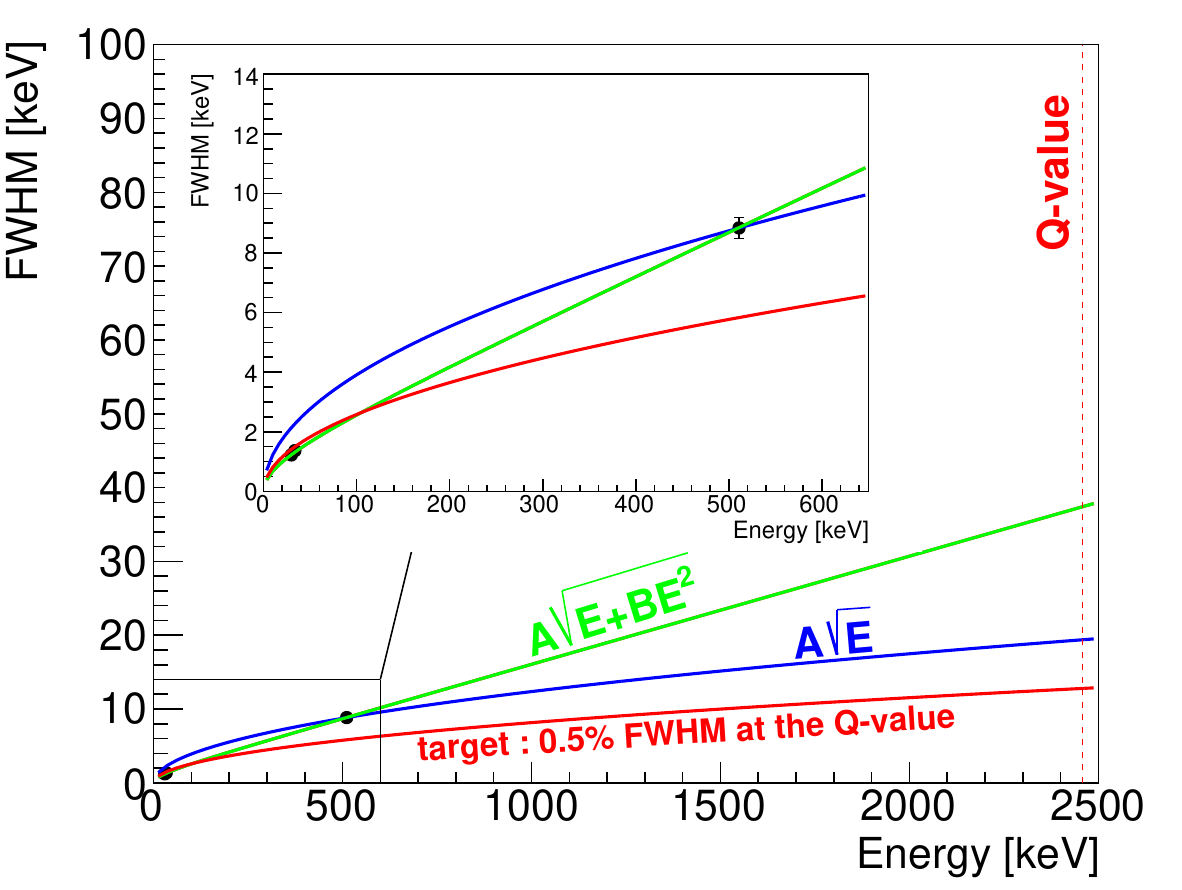}
        \caption{Extrapolation to the Q-value of ${}^{136}$Xe $0\nu\beta\beta$ decay with two types of function: $A\sqrt{E}$ and $A\sqrt{E+BE^2}$. The evaluation is performed with the resolution at 511~keV only for $A\sqrt{E}$ (blue curve) and with the resolutions at K$_\alpha$, K$_\beta$, and 511~keV peaks for $A\sqrt{E+BE^2}$ (green curve). The red curve represents our target energy resolution (0.5\% FWHM at the Q-value).}
        \label{fig:Analysis_extrapolation}
\end{figure}

Figure~\ref{fig:Analysis_extrapolation} shows the result. 
The extrapolated energy resolution (FWHM) at the Q-value, 2458~keV, is estimated to be 1.52\% (FWHM). 
This value does not reach the target resolution, 0.5\%, since the peak resolution at 511~keV is worse than the resolution of the characteristic X-ray peaks. The reason will be investigated by evaluating the expanded 180~L prototype using gamma-rays of higher energy in the future.
It is possible that the sensitive area is restricted by the cut in Section~\ref{sec:additional_z_cut}, and events with a small spread in the $z$~direction are collected selectively. Study with larger fiducial volume is desired.

\subsection{Event topology}
\begin{figure}[htb] 
	    \centering
    	\includegraphics[width=0.84\linewidth]{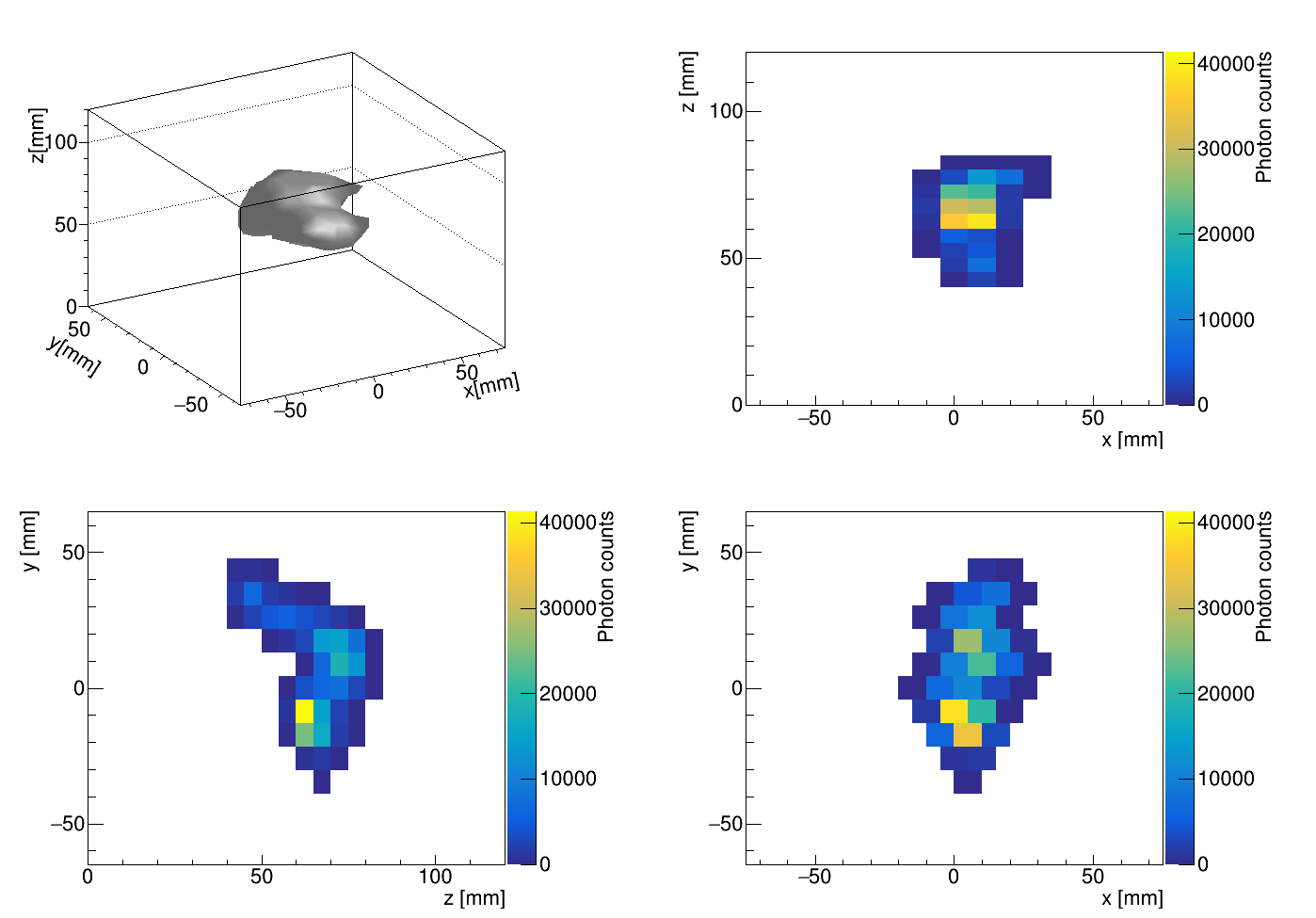} 
    	\caption{Event display of an event with 511~keV of deposited energy. The $z$ direction is sampled every 0.22~mm, but in this event display it is merged to 5~mm for easy viewing.}
    	\label{fig:event_display} 
\end{figure}

Figure~\ref{fig:event_display} shows an example event display of a 511~keV event. 
A blob structure at the track endpoint can be clearly seen. 
The number of blobs can be an index indicating the number of electron tracks. By determining the number of blobs, gamma-ray backgrounds and $0\nu\beta\beta$ signals can be distinguished. 
Eye scan shows that about one-half of the events have similar structures at their track endpoint. Five more event displays with the energy deposit of 511~keV are shown in~\ref{apx:event_display}.
Algorithms to distinguish $0\nu\beta\beta$ signals from gamma-ray backgrounds using topological information and machine learning method have been actively studied in xenon gas TPC experiments for $0\nu\beta\beta$ decay search~\cite{next_DL_2017}~\cite{pandaX_DL_2018}. We are also studying an algorithm based on DenseNet~\cite{densenet_2017} for future physics run~\cite{obara_VCI2019}.

\section{Conclusion}

AXEL is a high-pressure xenon gas TPC with a unique cellular readout scheme, ELCC,  
that is being designed to search for $0\nu\beta\beta$ decay. 
We developed a 180~L size prototype detector with excellent energy resolution and scalability. 
The dimension of the ELCC 
has been optimized  
using simulations
to achieve an energy resolution of 0.5\% at 
the ${}^{136}$Xe $0\nu\beta\beta$ Q-value, 2458~keV. 
Commissioning data was taken at 4~bar with 511~keV gamma-rays from a $^{22}$Na source and the obtained energy resolution is 1.73 $\pm$ 0.07\% (FWHM).
The energy resolution at 2458~keV was estimated to be 0.79~$\pm$~0.03\% (FWHM) based on extrapolation from only the 511~keV peak using a $A\sqrt{E}$ function.
Combining with evaluations of the K$_\alpha$ and K$_\beta$ peaks, the estimated energy resolution at the Q-value is 0.79--1.52\% (FWHM).
Ionization electron tracks are reconstructed from the hit patterns and hit timings in the ELCC. The structure at the end point of electron track (blob) can be seen in these track information. Measurement at higher energy will be performed with the upgraded next phase detector and higher pressure to further improve the energy resolution and demonstrate the performance at the $0\nu\beta\beta$ Q-value.

\section*{Acknowledgment}
The AXEL experiment is supported 
by the JSPS KAKENHI Grant Numbers
15H02088, 
16J09462, 
17K18777, 
17J00268, 
18J13957, 
18J00365, 
18J20453, 
and 18H05540.
We also thank the Institute for Cosmic Ray Research and Kamioka underground laboratory, the University of Tokyo, for their support of our project.
The development of the electronics board is supported by Open-It (Open Source Consortium of Instrumentation), High Energy Accelerator Research Organization (KEK).
We thank R.~Wendell for his help preparing the manuscript.

\bibliographystyle{myptephy}


\appendix
\def\thesection{Appendix \Alph{section}}

\section{MPPC recovery time}
\label{apx:MPPCrecovery}
To model the relationship between the output signal of the MPPC and the number of incident photons, we consider 
$N_{\textrm{observed}}$ photons incident on a MPPC during $\Delta t$ seconds. 
Here, only photons which create an electron-hole pair in the MPPC sensitive region are considered.
Using the amount of photons per unit time per MPPC pixel, $k \equiv N_{\textrm{observed}}/(\Delta t\cdot N_{\textrm{pixel}})$, where $N_{\textrm{pixel}} = 3600$ is the number of pixels on a S17330 MPPC and $\Delta t$ is set to 200~ns corresponding to 1~clock of the ADC, the probability that a photon enters 
a particular MPPC pixel again $t$-seconds after the pixel detects 
a proceeding can be expressed as $ke^{kt}$.

The recovery time $\tau$ is defined as the time required for the pixel gain to recover to $(1-1/e)$ times the original gain, $g_0$, after the pixel emits a pulse. 
Thus, a gain $t$-seconds after the previous pulse is represented as $g_0 (1-e^{-t/ \tau})$, and the average gain $g$ is calculated as
\begin{equation} g = \int^{\infty}_{0} ke^{kt} g_0(1-e^{-\frac{t}{\tau}}) dt = \frac{g_0}{1+k\tau}. \end{equation}
The output by a MPPC with gain $g$ is
\begin{equation} N'_{\textrm{observed}} = \frac{N_{\textrm{observed}}}{1+k\tau}. \label{eq:N_obs}\end{equation}
Hence, solving for $N_{\textrm{observed}}$, the number of true incident photons can be estimated with the recovery time $\tau$ from the number of observed photons $N'_{\textrm{observed}}$ as in Equation~(\ref{eq:MPPCsatu}).

\section{Additional event display}
\label{apx:event_display}
\begin{figure}[htb] 
	    \centering
    	\includegraphics[width=0.9\linewidth]{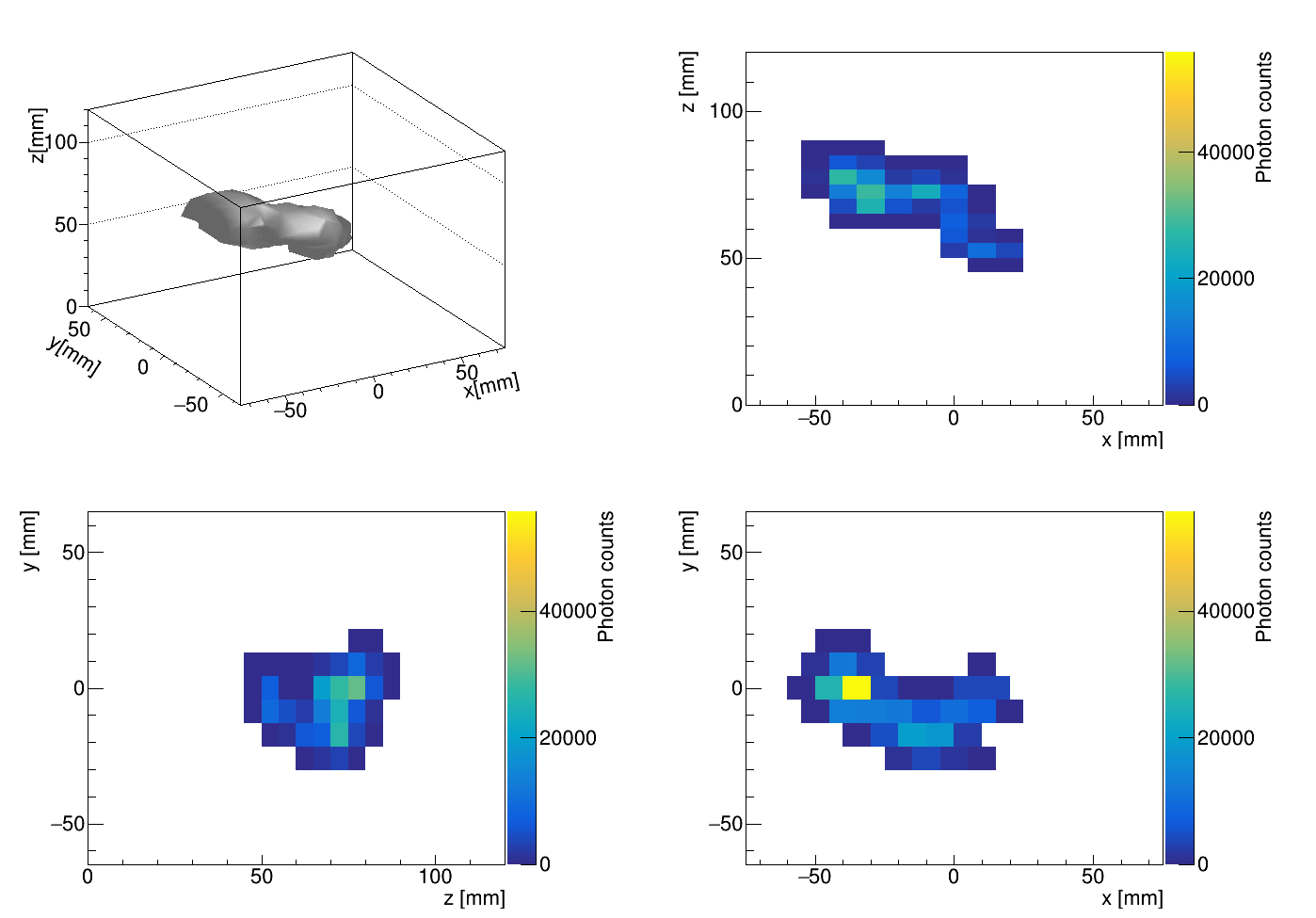} 
    	\caption{Event display of a 511~keV event}
    	\label{fig:event_display2} 
\end{figure}

\begin{figure}[htb] 
	    \centering
    	\includegraphics[width=0.9\linewidth]{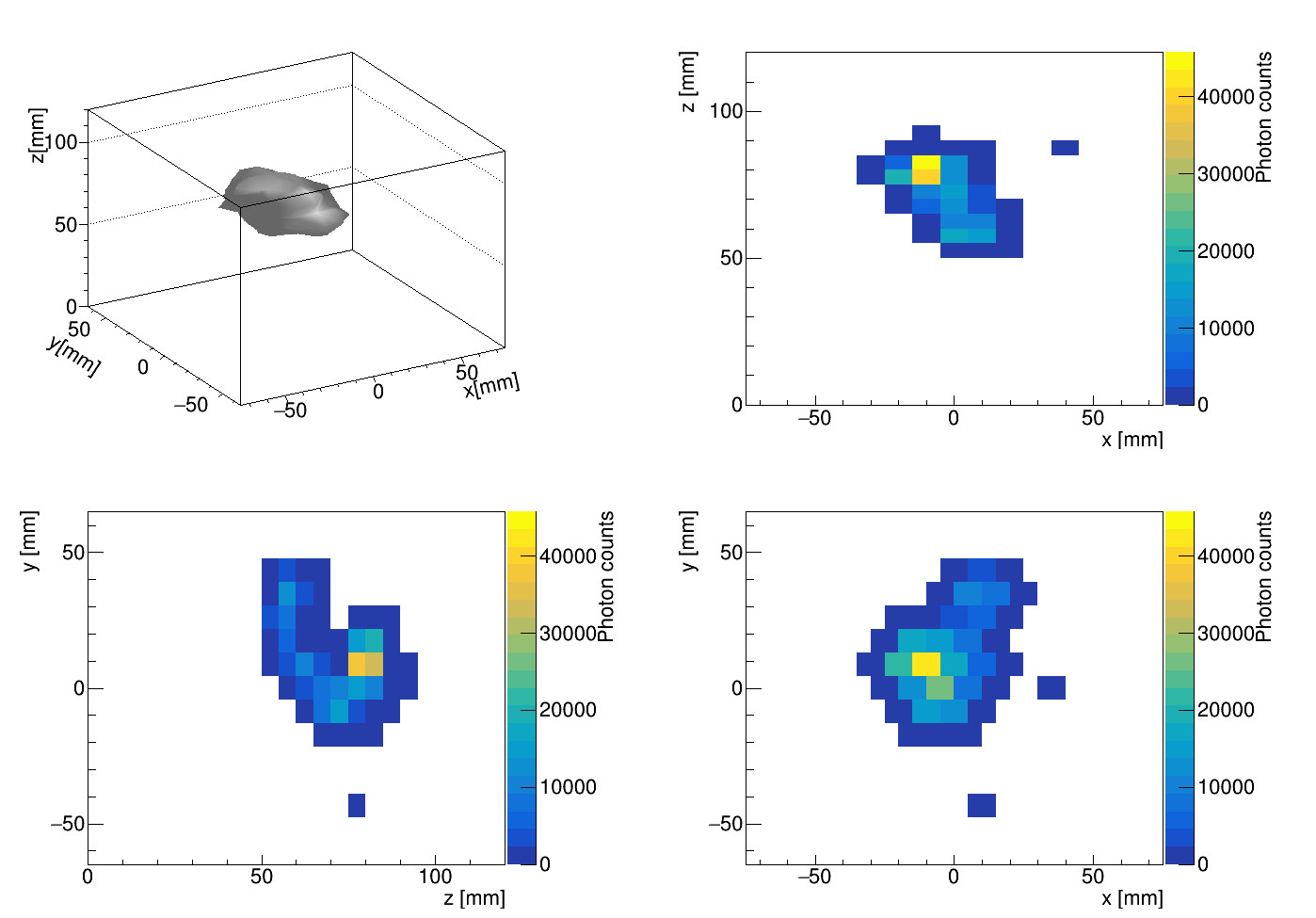} 
    	\caption{Event display of a 511~keV event}
    	\label{fig:event_display3} 
\end{figure}

\begin{figure}[htb] 
	    \centering
    	\includegraphics[width=0.9\linewidth]{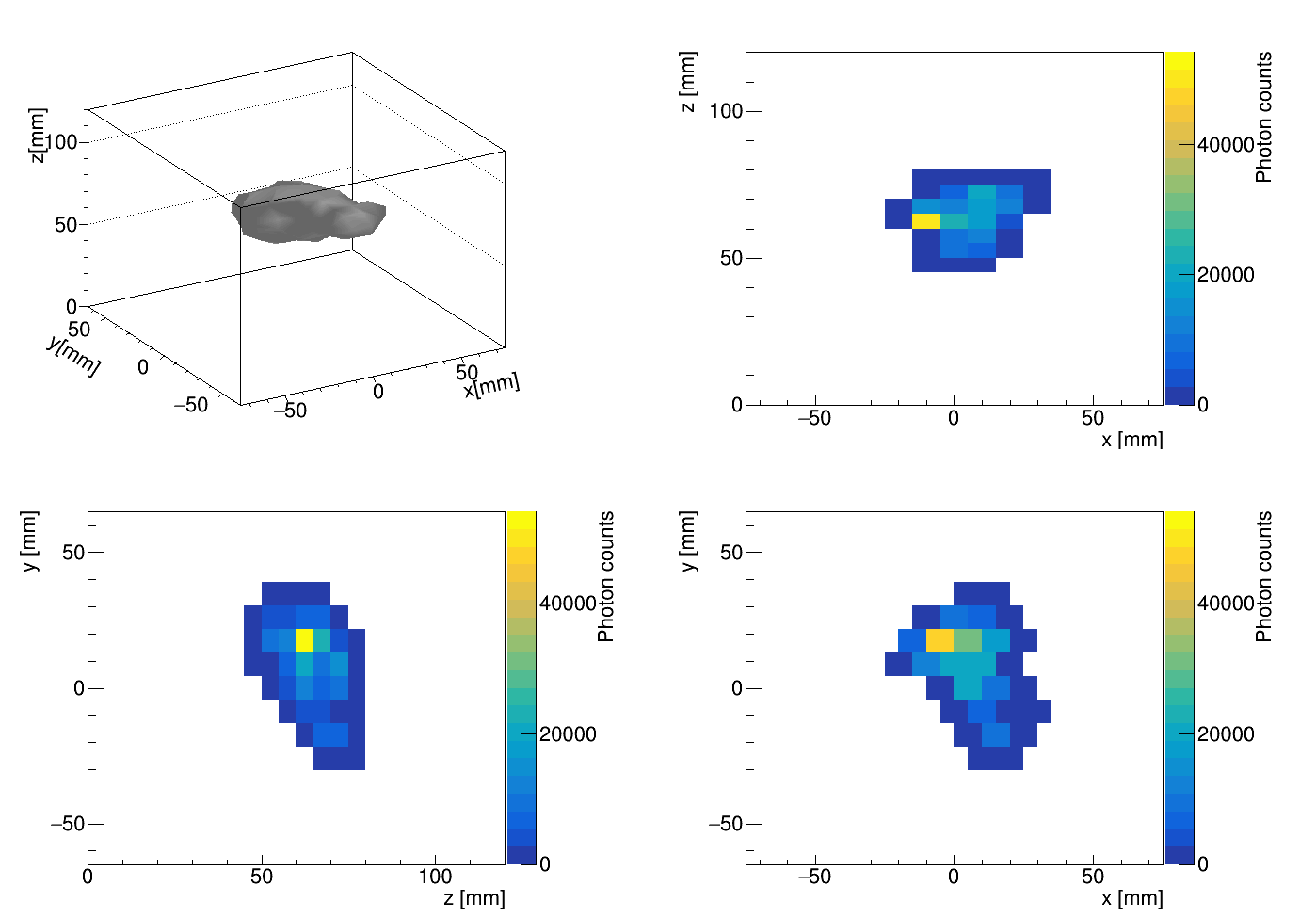} 
    	\caption{Event display of a 511~keV event}
    	\label{fig:event_display4} 
\end{figure}

\begin{figure}[htb] 
	    \centering
    	\includegraphics[width=0.9\linewidth]{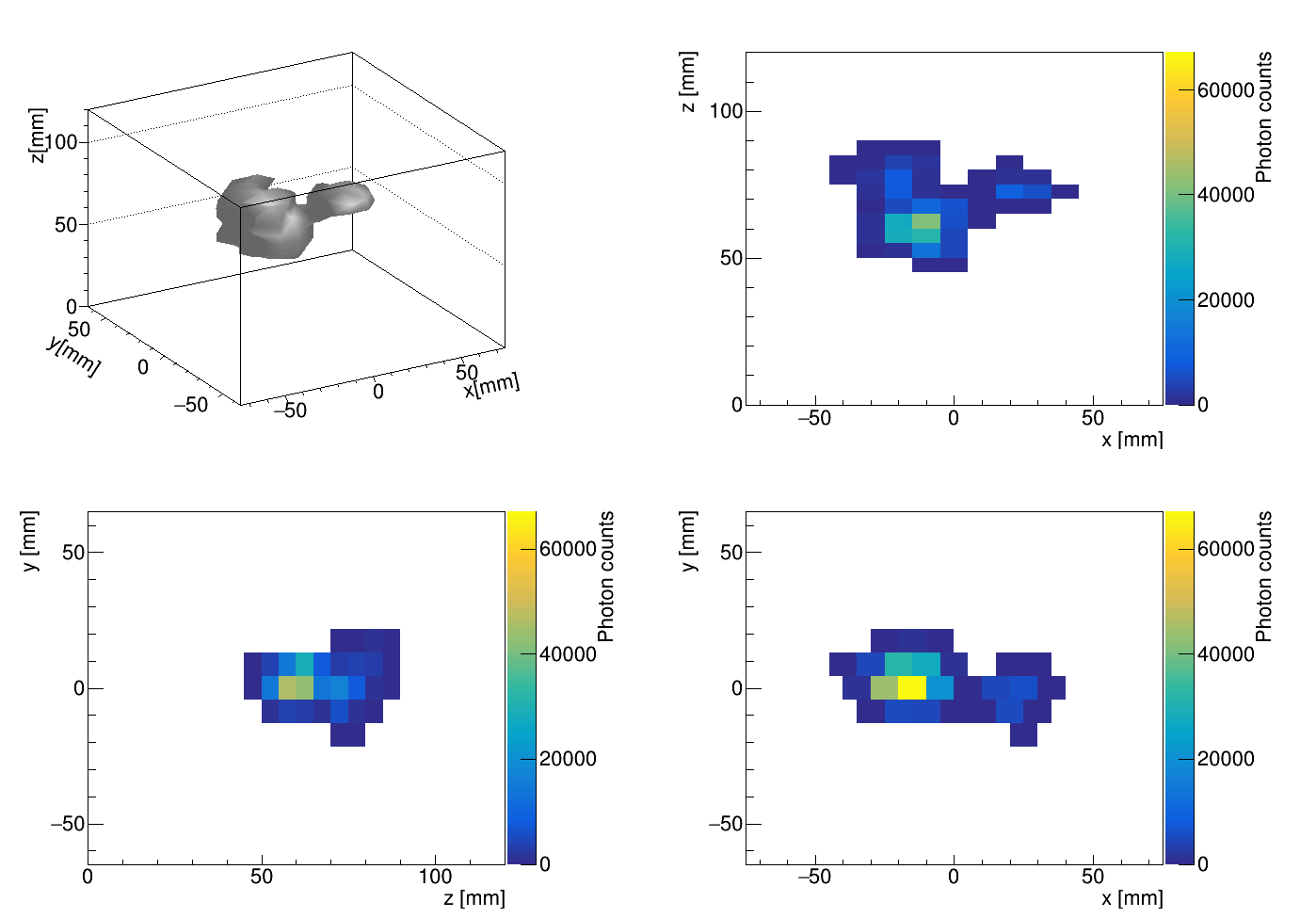} 
    	\caption{Event display of a 511~keV event}
    	\label{fig:event_display5} 
\end{figure}

\begin{figure}[htb] 
	    \centering
    	\includegraphics[width=0.9\linewidth]{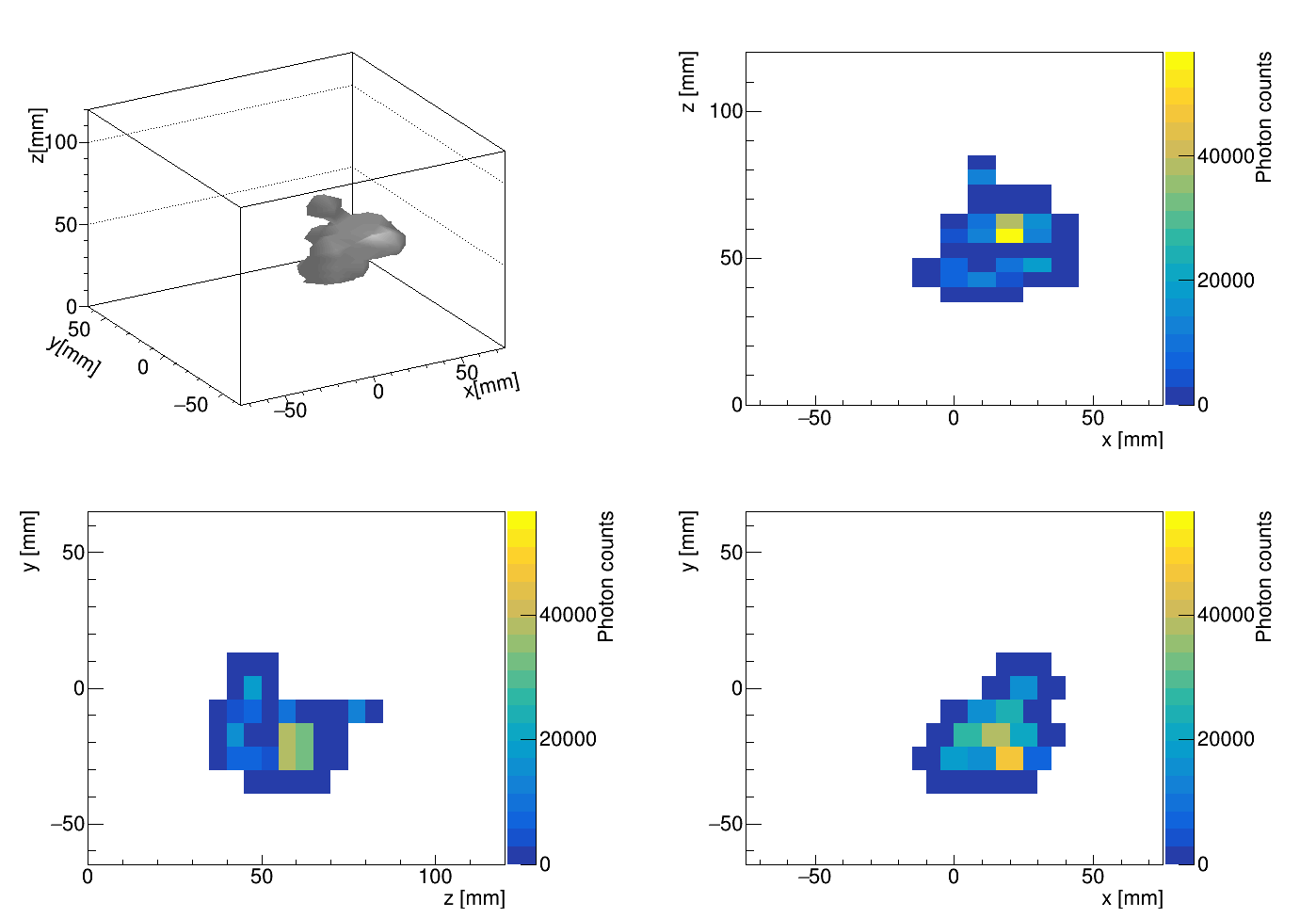} 
    	\caption{Event display of a 511~keV event}
    	\label{fig:event_display6} 
\end{figure}

Five example event displays with the deposited of 511~keV are shown in Figures~\ref{fig:event_display2}--\ref{fig:event_display6}.
In these figures, data along $z$ direction is merged to 5~mm as described in Figure~\ref{fig:event_display}.

\end{document}